\documentclass[a4paper]{jpconf}
\usepackage{graphicx}
\usepackage{amsmath,amssymb}
\usepackage{soul,color}
\usepackage[dvipsnames]{xcolor}
\begin{document}
\title{Multiplicative Anomaly matches Casimir Energy for GJMS Operators on Spheres}

\author{R Aros$^1$, F Bugini$^2$, DE D\'iaz$^3$ and B Z\'u\~niga$^1$}

\address{$^1$ Departamento de Ciencias Fisicas, Universidad Andres Bello,
Sazie 2212, Piso 7, Santiago, Chile}
\address{$^2$ Departamento de Matem\'atica y F\'isica Aplicadas,
Universidad Cat\'olica de la Sant\'isima Concepci\'on,
Alonso de Ribera 2850, Concepci\'on, Chile}
\address{$^3$ Departamento de Ciencias Fisicas, Universidad Andres Bello,
Autopista Concepcion-Talcahuano 7100, Talcahuano, Chile}
\ead{raros@unab.cl,bugini@ucsc.cl,danilodiaz@unab.cl,b.zuigacontreras@uandresbello.edu}
\begin{abstract}
An explicit formula to compute the multiplicative anomaly or defect of $\zeta$-regularized products of linear factors is derived, by using a Feynman parametrization, generalizing Shintani-Mizuno formulas. Firstly, this is applied on $n$-spheres, reproducing known results in the literature. Then, this framework is applied to a closed Einstein universe at finite temperature, namely $S^1_{\beta}\times S^{n-1}$. In doing so, it is shown that the standard Casimir energy (as computed via $\zeta$ regularization)  for GJMS operators coincides with the accumulated multiplicative anomaly for the shifted Laplacians that build them up. This equivalence between Casimir energy and multiplicative anomaly within $\zeta$ regularization, unnoticed so far to our knowledge, brings about a new turn regarding the physical significance of the multiplicative anomaly, putting both now on equal footing. An emergent {\it improved} Casimir energy, that incorporates the multiplicative anomaly among the building Laplacians, is also discussed.
\end{abstract}

\section{Introduction}

It has long been known that $\zeta$-regularized functional determinants of differential operators \cite{RAY1971145} may be afflicted by a {\it multiplicative anomaly} \cite{Kontsevich1995}.
Even for commuting (elliptic) differential operators A and B, in general
$\det_{\zeta}{\left(A\cdot B\right)}\neq \det_{\zeta}{A}\cdot \det_{\zeta}{B}$. In the mid-80s, an explicit expression for this multiplicative anomaly was devised by Wodzicki in terms of the so-called non-commutative residue \cite{wodzicki1987noncommutative,Wodzicki1987,GUILLEMIN1985131}.  Interestingly, in cases where the eigenvalues factorize into linear factors the discrepancy for the resulting $\zeta$-regularized products had been pinpointed a decade before in Shintani's works~\cite{ShintaniOnValues,10.3836/tjm/1270472992}. In these cases, the individual $\zeta$ functions are Barnes multiple zetas~\cite{barnes1904theory} and the collective ones are the Shintani-Barnes generalizations thereof~\cite{Friedman2004362, MIZUNO2006155}.
The direct connection between the two approaches is enabled by a crucial feature of the $\zeta$-regularized products: the multiplicative anomaly between several factors is {\it pairwise accumulative}, i.e. it is enough to compute it between all possible pairings and then average the result
~\cite{CastilloFriedman2011,CastilloFriedman2012,Dowker:2014tea}. Therefore, although the multiplicative anomaly between linear factors may not be captured by Wodzicki's formula, the converse holds: the multiplicative anomaly between the quadratic factors in the Laplacians, for which Wodzicki's formula applies, is equally captured by the multiplicative anomaly among the linear factors. Another remarkable feature of the multiplicative anomaly between linear factors is that, as compared to the $\zeta$-regularized products that involve Shintani-Barnes gammas, it is far simpler. In all known examples it reduces to an exponential of a rational function in the coefficients of the linear factors and the logarithms of these coefficients (see, \textit{e.g.} \cite{ShintaniOnValues,Friedman2004362,MIZUNO2006155,CastilloFriedman2011}).

Motivated by these results, in this note, we revisit the computation of Casimir or vacuum energy for higher-derivative operators on spheres since the standard calculation seems to overlook the possible multiplicative anomaly among the factors (see, e.g.,~\cite{Beccaria:2014xda} for the Paneitz operator in 4D). For concreteness, we focus on conformal powers of the Laplacian or GJMS operators $P_{2k}$~\cite{GJMSOriginal1992} which happen to factorize into shifted Laplacians on spheres $S^n$ \cite{BransonSharp1995, Gover2005}, as well as on the conformally flat product space $S^1_{\beta}\times S^{n-1}$ \cite{juhl2010conformally, Beccaria_2016}. The partition function on the latter geometry is dominated by the Casimir energy in the low-temperature ($\beta\rightarrow\infty$) limit, and the presence of a multiplicative anomaly leads to an {\it improved} Casimir energy. The improvement relies on consistency: there are two alternative factorizations in terms of shifted Laplacian and only after the inclusion of the multiplicative anomaly one can find agreement between the partition functions and, in consequence, between the Casimir energies. In addition, on the two torus, the universal dependence of the Casimir energy on the central charge $E_0=-c/12$ is restored.

The organization of this paper is as follows. In Section 2, a generalization of Sintani-Mizuno formulas for the multiplicative anomaly of linear factors is obtained by a procedure based on Feynman parametrization and Fock-Schwinger-DeWitt proper-time representation. In Section 3, Mizuno's result in two dimensions is extended to three and four dimensions, casting the answer into Bernoulli polynomials and keeping the quasi-periods in the greatest generality. Since the explicit expressions become increasingly cluttered as the dimensions grow, we restrict attention to particular examples in what follows. In section 4, as a preliminary exercise, the multiplicative anomaly is computed on round spheres confirming previous results in the literature. Then, in Section 5, the background of present interest is addressed, namely the closed Einstein universe at finite temperature, and new features of the Casimir energy are reported. In Section 6, we highlight the role of the multiplicative anomaly in Shintani's derivation of the Kronecker second limit formula. In Section 7, as the main application, we examine the computation of the Casimir energy for GJMS operators in the light again of the new features that the inclusion of the multiplicative anomaly brings in. Summary and outlook are provided in Section 8. Finally, miscellaneous results are collected in two appendices.

\section{Derivation of Shintani-Mizuno formulas via Feynman parametrization}

To compute the ratio of the functional determinants, and of the corresponding $\zeta-$regularized products, we start with the relative zeta function
\begin{eqnarray}
\zeta_{AB}(s)-\zeta_{A}(s)-\zeta_{B}(s)=\sum_{\vec{m}=\vec{0}}^{\vec{\infty}}\left\{\frac{1}{[\vec{m}\cdot\vec{a}+w]^s\cdot[\vec{m}\cdot\vec{b}+z]^s}-\frac{1}{[\vec{m}\cdot\vec{a}+w]^s}-
\frac{1}{[\vec{m}\cdot\vec{b}+z]^s}\right\}
\end{eqnarray}
for $\Re{(s)}>n$, with the multi-index $\vec{m}=(m_1,m_2,...,m_n)$ being an $n$-tuple of non-negative integers and assuming that the quasi-periods $a_i$ and $b_i$, as well as the arguments $w$ and $z$, all have positive real parts (although this may be relaxed later, as we will see).
We now combine the factors in the first term into a single denominator by using Feynman parametrization while the second and third terms come in for free
\begin{eqnarray}
\frac{\Gamma(2s)}{\Gamma^2(s)}\int_0^1dv\,[v\,(1-v)]^{s-1}\left\{ \frac{1}{[\vec{m}\cdot(\vec{a}\,v + (1-v)\,\vec{b})+w\,v+(1-v)\,z]^{2s}}\right.\nonumber\\
-\left.\frac{1}{[\vec{m}\cdot\vec{a}+w]^{s}}
-\frac{1}{[\vec{m}\cdot\vec{b}+z]^{s}}
\right\}
\end{eqnarray}
Next, we introduce Fock-Schwinger-DeWitt proper-time representations for the inverse powers
\begin{eqnarray}
 && \frac{1}{\Gamma(2s)}\int_0^{\infty}\frac{dt}{t}\,t^{2s}\,e^{-t[\vec{m}\cdot(\vec{a}\,v + (1-v)\,\vec{b})+w\,v+(1-v)\,z]}\nonumber \\
&-&\frac{1}{\Gamma(s)}\left(\int_0^{\infty}\frac{dt}{t}\,t^{s}\,e^{-t[\vec{m}\cdot\vec{a}+w]} + \int_0^{\infty}\frac{dt}{t}\,t^{s}\,e^{-t[\vec{m}\cdot\vec{b}+z]}\right)
\end{eqnarray}
The geometric series summation in the multi-index
$\vec{m}$ produces the following Bose factors
\begin{eqnarray}
\nonumber
&&\frac{1}{\Gamma(2s)}\int_0^{\infty}\frac{dt}{t}\,t^{2s}\,\frac{e^{-t[w\,v+(1-v)\,z]}}{\prod_{i=1}^n\left\{1-e^{-t(a_i\,v + (1-v)\,b_i)}\right\}}\qquad\qquad\qquad\qquad\\\nonumber
\\
&-&\frac{1}{\Gamma(s)}\int_0^{\infty}\frac{dt}{t}\,t^{s}\,\frac{e^{-t\,w}}{\prod_{i=1}^n\left\{1-e^{-t\,a_i }\right\}}
-\frac{1}{\Gamma(s)}\int_0^{\infty}\frac{dt}{t}\,t^{s}\,\frac{e^{-t\,z}}{\prod_{i=1}^n\left\{1-e^{-t\,b_i }\right\}}
\end{eqnarray}
The Bose factors are now expressed as a Taylor series in $t$ with Bernoulli numbers\footnote{We use the convention $B^+_k= B_k(1)$, as opposed to $B^-_k= B_k(0)$, in terms of the Bernoulli polynomials.} as coefficients
\begin{eqnarray}
    && \sum_{\vec{l}=\vec{0}}^{\vec{\infty}} \left\{\prod_{i=1}^{n}\frac{B^+_{l_i}}{(l_i)!}\,(a_i\,v + (1-v)\,b_i)^{l_i-1}\right\} \left\{ \frac{1}{\Gamma(2s)}\int_0^{\infty}\frac{dt}{t}\,t^{2s-n+\sum_{i=1}^n l_i} e^{-t[w\,v+(1-v)\,z]}\right\} \nonumber\\
    &-& \sum_{\vec{l}=\vec{0}}^{\vec{\infty}} \left\{\prod_{i=1}^{n}\frac{B^+_{l_i}}{(l_i)!}\,a_i^{l_i-1}\right\} \left\{\frac{1}{\Gamma(s)}\int_{0}^{\infty}\frac{dt}{t}\,t^{s-n+\sum_{i=1}^n l_i}\,e^{-t\,w}\right\} \\
    &-& \sum_{\vec{l}=\vec{0}}^{\vec{\infty}}\,\left\{\prod_{i=1}^{n}\frac{B^+_{l_i}}{(l_i)!}\,b_i^{l_i-1}\right\} \left\{\frac{1}{\Gamma(s)}\int_0^{\infty}\frac{dt}{t}\,t^{s-n+\sum_{i=1}^n l_i}\,e^{-t\,z}\right\} \nonumber
\end{eqnarray}
The proper-time integrals, taken in terms of Euler gamma functions, enable the analytic continuation in the spectral parameter $s$, and yield\begin{eqnarray}
    && \sum_{\vec{l}=\vec{0}}^{\vec{\infty}}\,\left\{\prod_{i=1}^{n}\frac{B^+_{l_i}}{(l_i)!}\,(a_i\,v + (1-v)\,b_i)^{l_i-1}\right\}\frac{\Gamma(2s-n+\sum_{i=1}^n l_i)}{\Gamma(2s)}\,\left[w\,v+(1-v)\,z\right]^{-2s+n-\sum_{i=1}^n l_i} \nonumber\\
    &-& \sum_{\vec{l}=\vec{0}}^{\vec{\infty}}\,\left\{\prod_{i=1}^{n}\frac{B^+_{l_i}}{(l_i)!}\,a_i^{l_i-1}\right\}\frac{\Gamma(s-n+\sum_{i=1}^n l_i)}{\Gamma(s)}\,\left[w\right]^{-s+n-\sum_{i=1}^n l_i}\\
    &-& \sum_{\vec{l}=\vec{0}}^{\vec{\infty}}\,\left\{\prod_{i=1}^{n}\frac{B^+_{l_i}}{(l_i)!}\,b_i^{l_i-1}\right\}\frac{\Gamma(s-n+\sum_{i=1}^n l_i)}{\Gamma(s)}\,\left[z\right]^{-s+n-\sum_{i=1}^n l_i} \nonumber
\end{eqnarray}
The $\zeta$-regularized products are obtained from the derivative of the $\zeta$ with respect to the spectral parameter $s$ at $s=0$. By careful examination of the behavior as $s\rightarrow 0$ one can realize 
\begin{eqnarray}
  \zeta_{AB}(s)-\zeta_{A}(s)-\zeta_{B}(s) =\frac{s}{2}\times \mbox{Regular}+\frac{1}{2}\zeta_{A}(2s)+\frac{1}{2}\zeta_{B}(2s)-\zeta_{A}(s)-\zeta_{B}(s)+ O(s^3)~.
\end{eqnarray}
As a consistency check, direct evaluation at $s=0$ results in the additive property of the zeta function weighted by the order,  2 for AB and 1 for A and for B, of the corresponding differential operators  
\begin{equation}
\zeta_{AB}(0)=\frac{1}{2}\,\zeta_{A}(0)+\frac{1}{2}\,\zeta_{B}(0)~.
\end{equation}
Back to the regularized products, the overall prefactor $\Gamma(2s)/\Gamma^2(s)$ goes as $s/2$ and the rest is regular at $s=0$, after symmetrization with respect to $v\leftrightarrow 1-v$, so it is enough to consider the limit of the latter as $s\rightarrow0$ to compute the derivative at zero.
In addition, the factors $\Gamma(2s-n+\sum_{i=1}^n l_i)/\Gamma(2s)$ and $\Gamma(s-n+\sum_{i=1}^n l_i)/\Gamma(s)$ in the limit $s\rightarrow0$ produce a vanishing result unless the numerators also hit a pole, say $-p$ with $p=0,1,2,...,n$, and gives a finite answer  $(-1)^p/p!$. Therefore the sums over $l_i\geq0$ are truncated by the condition $p+\sum_{i=1}^n l_i=n$
\begin{eqnarray}
&& -\frac{1}{2}\sum_{l_i,p\geq 0 }\,\left\{\frac{(-1)^p}{p!}\prod_{i=1}^{n}\frac{B^+_{l_i}}{(l_i)!}\right\}\,\int_0^{1}\frac{dv}{v(1-v)} \\\nonumber
&\times&\frac{1}{2}\left\{\left\{\prod_{i=1}^{n}(a_i\,v + (1-v)\,b_i)^{l_i-1}\right\}\,\left[w\,v+(1-v)\,z\right]^{p} +\left\{(\vec{a},w)\leftrightarrow(\vec{b},z)\right\}\right.
\\\nonumber&&\left.-\left\{\prod_{i=1}^{n}a_i^{l_i-1}\right\}\,[w]^{p}-\left\{\prod_{i=1}^{n}b_i^{l_i-1}\right\}\,[z]^{p}\right\} \nonumber
\end{eqnarray}
One last change of variables in the Feynman parameter $1/v-1=u$ and realizing that the inversion $u\rightarrow1/u$ merely interchanges $(\vec{a},w)\leftrightarrow(\vec{b},z)$, cast the final result in Shintani-Mizuno~\cite{ShintaniOnValues,MIZUNO2006155} form
\begin{eqnarray}
  \mbox{MA}(A,B) &=&-\zeta'_{AB}(0)+\zeta'_{A}(0)+\zeta'_{B}(0) \\
  &=&\left.-\frac{1}{2}\sum_{l_i,p\geq 0}\,\left\{\frac{(-1)^p}{p!}\prod_{i=1}^{n}\frac{B^+_{l_i}}{(l_i)!}\right\}C(\vec{a},z;\vec{b},w\,|\,\vec{l},p)\right|_{p +\sum_{i=1}^n l_i=n}\nonumber,
\end{eqnarray}
with
\begin{eqnarray}
C(\vec{a},w;\vec{b},z\,|\,\vec{l},p) &=&\int_0^1\,\frac{du}{u}\left\{\left\{\prod_{i=1}^{n}(a_i + u\,b_i)^{l_i-1}\right\}\,\left[w+u\,z\right]^{p}
-\left\{\prod_{i=1}^{n}a_i^{l_i-1}\right\}\,[w]^{p}\right\}  \nonumber\\ &+& \left\{(\vec{a},w)\leftrightarrow(\vec{b},z)\right\}.
\end{eqnarray}
This formula, which computes the multiplicative anomaly for a pair of linear factors, suffices to deal with a generic number of linear factors because, as already mentioned, the multiplicative anomaly turns out to be pairwise accumulative~\cite{CastilloFriedman2011, CastilloFriedman2012,Dowker:2014tea}.\\
The $n=2$ case corresponds to the formula put forward by Mizuno~(cf. proof of Lemma 4 in~\cite{MIZUNO2006155}), whereas the original Shintani formula (cf. Proposition 1 in~\cite{ShintaniOnValues}) applies to the particular choice of the arguments $w=\vec{a}\cdot\vec{x}$ and $z=\vec{b}\cdot\vec{x}$. In that case, the polynomial dependence on $x_i$ can be rearranged by trading the Bernoulli numbers by Bernoulli polynomials in $x_i$ after summing over $p$. This is easily seen by going back to the Bose factors and expanding each of them in terms of Bernoulli polynomials in $x_i$. \\
One can alternatively choose to expand the whole product of Bose factors in terms of Bernoulli polynomials of higher order\footnote{Our convention differs slightly from the definition in \cite{ErdelyBateman} or \cite{RUIJSENAARS2000107}.},
\begin{equation}
\frac{t^n\,e^{-w\,t}}{\prod_{i=1}^n\left\{1-e^{-a_i\,t}\right\}}=\sum_{l=0}^{\infty}B_{n,l}(w| \vec{a})\,\frac{t^l}{l!},
\end{equation}
in which case the integral formula for the multiplicative anomaly can be written in the following neat and compact form
\begin{equation}
  \mbox{MA}(A,B) =
-\frac{1}{2\,n!}\int_0^1\,\frac{du}{u}\left\{B_{n,n}(w+u\,z \,|\,\vec{a}+u\,\vec{b})-B_{n,n}(w \,|\,\vec{a}) \right\}+  \left\{(\vec{a},w)\leftrightarrow(\vec{b},z)\right\}.
\end{equation}
Moreover, in our convention, the Bernoulli polynomial of higher order coincides essentially with Barnes zeta at $s=0$
\begin{equation}
\frac{B_{n,n}(w,\vec{a})}{n!} = \zeta_n(0,w|\vec{a})~.
\end{equation}
Consequently, the multiplicative anomaly becomes an average of Barnes $\zeta$'s with respect to the Feynman parameter $u$
\begin{equation}
\boxed{\mbox{MA}(A,B) =
-\frac{1}{2}\int_0^1\,\frac{du}{u}\left\{\zeta_{n}(0,w+u\,z \,|\,\vec{a}+u\,\vec{b})-\zeta_{n}(0,w \,|\,\vec{a}) \right\}+  \left\{(\vec{a},w)\leftrightarrow(\vec{b},z)\right\}.}
\end{equation}

\section{Previous and new results}
Let us now put the formula to work, keeping the quasi-periods in the greatest generality.
\subsection{1D: Friedman-Ruijsenaars formula}
The $n=1$ case was worked out by Friedman and Ruijsenaars~\cite{Friedman2004362} some time ago by exploiting a recurrence relation and, as expected, their result matches the outcome of the Shintani-Mizuno integral representation above
\begin{equation}
\mbox{MA}(A,B)=\frac{1}{2}\cdot\left(\frac{w}{a}-\frac{z}{b}\right)\cdot \log{\frac{a}{b}}
\end{equation}
\subsection{2D: Shintani-Mizuno formula}
The $n=2$ case was addressed by Mizuno~\cite{MIZUNO2006155}, following Shintani's approach, and his result was concisely written in terms of the Bernoulli polynomial of order 2 as follows
\begin{equation}
\mbox{MA}(A,B)=\frac{a_1\,b_2\,-\,a_2\,b_1}{4\,a_1\,b_1}\cdot B_2\left(\frac{a_1\,z\,-\,b_1\,w}{a_1\,b_2\,-\,b_1\,a_2}\right)\cdot\log\frac{a_1}{b_1}\;+\;\{1\leftrightarrow2\}
\end{equation}

\subsection{3D: generalized Shintani-Mizuno formula}
We report here the $n=3$ case for the first time, to our knowledge, obtained with MAPLE help to compute the integrals and to concisely express the result in terms of Bernoulli polynomials
\begin{eqnarray}
\mbox{MA}(A,B)&=&-\left\{\frac{[a_1\,b_2\,-\,a_2\,b_1\,+\,a_1\,b_3\,-\,a_3\,b_1]^3}{12\,a_1\,b_1\,(a_1\,b_2\,-\,a_2\,b_1)(a_1\,b_3\,-\,a_3\,b_1)}\cdot B_3\left(\frac{a_1\,z\,-\,b_1\,w}{a_1\,b_2\,-\,a_2\,b_1\,+\,a_1\,b_3\,-\,a_3\,b_1}\right)\right.\nonumber\\ & &\nonumber\\
&+& \left.\frac{[a_1\,b_2\,-\,a_2\,b_1\,+\,a_1\,b_3\,-\,a_3\,b_1]}{24\,a_1\,b_1}\cdot B_1\left(\frac{a_1\,z\,-\,b_1\,w
}{a_1\,b_2\,-\,a_2\,b_1\,+\,a_1\,b_3\,-\,a_3\,b_1}\right)
\right\}\cdot\log\frac{a_1}{b_1} \nonumber\\ & &\nonumber\\
&-& \{1\leftrightarrow 2\} - \{1\leftrightarrow 3\} \label{MAB2}.
\end{eqnarray}

\subsection{4D: generalized Shintani-Mizuno formula}
For $n=4$ the answer, also new to our knowledge, becomes more involved. We introduce some notation to write it down more compactly:
\begin{eqnarray}
D_{ij}&=&a_ib_j-b_ia
_j \nonumber\\
D&=&D_{12}+D_{13}+D_{14}
\end{eqnarray}
\begin{eqnarray}
\mbox{MA}(A,B)&=& \left\{\frac{D^4}{48\,a_1\,b_1\,D_{12}\,D_{13}\,D_{14}}\cdot B_4\left(\frac{a_1\,z\,-\,b_1\,w}{D}\right)\right.\nonumber\\ & &\nonumber\\
&-& \left.\frac{D^2\,(D_{12}^2+D_{13}^2+D_{14}^2-D^2)}{96\,a_1\,b_1\,D_{13}\,D_{14}\,D_{12}}\cdot B_2
\left(\frac{a_1\,z\,-\,b_1\,w
}{D}\right)\right.\nonumber\\ & &\nonumber\\
&-& \frac{D^4-D^2\,(D_{12}^2+D_{13}^2+D_{14}^2)-2 DD_{12}D_{13}D_{14}}{2880\,a_1\,b_1\,D_{12}\,D_{13}\,D_{14}} \nonumber\\
& & \nonumber\\
&+& \left. \frac{3(D_{12}D_{13}+D_{12}D_{14}+D_{13}D_{14})^2}{1440\,a_1\,b_1\,D_{12}\,D_{13}\,D_{14}} \right\} \cdot\log\frac{a_1}{b_1}  \nonumber\\
& & \nonumber\\
&+& \{1\leftrightarrow 2\} + \{1\leftrightarrow 3\} + \{1\leftrightarrow 4\} \label{MAB3}.
\end{eqnarray}
Since the explicit answer becomes more complicated as we increase the dimension, we refrain from displaying it for higher dimensions, and, in what follows, we focus on particular choices for the quasi-periods.

\section{Examples: shifted Laplacian on round spheres $S^n$}
Let us consider the factorization of the eigenvalues of the Laplacian, on (unit) spheres, shifted by a constant.
First, recall the eigenvalues and multiplicities for the (negative) Laplacian $-\nabla^2$ on the $n$-sphere $S^n$
\begin{equation}
    \lambda_l=l(l+n-1)\qquad\qquad \mbox{deg}(l)=(2l+n-1)\frac{(l+n-2)!}{l!\,(n-1)!}~.
\end{equation}
Notice that a shift by $\frac{(n-1)^2}{4}-a^2$ factorizes the eigenvalues $\Lambda_l$ of the shifted Laplacian $L_a=-\nabla^2+\frac{(n-1)^2}{4}-a^2=D_a\,D_{-a}$ into linear factors on $l$
\begin{equation}
    \Lambda_l=(l+\frac{n-1}{2}+a)(l+\frac{n-1}{2}-a)~.
\end{equation}
There is now a gracious way to connect with the regularized product of the previous sections: we follow Dowker's `central tactic' (see, e.g.~\cite{Dowker:2014tea,Dowkercmp/1104270298}) in the spectral treatment of the Laplacian on spheres which consists in taking the full sphere as the union of the Neumann and Dirichlet problems on the hemisphere. We trade then the `orbital' quantum number $l$ by the sum of non-negative integers $m_1+m_2+...+m_n$ for Neumann boundary condition and $1+m_1+m_2+...+m_n$ for Dirichlet. At fixed $l$ the combinatorics produce the correct multiplicity: for Neumann, the counting consists of the different ways to distribute $l$ balls in $n$ boxes,
whereas for Dirichlet there are only $l-1$ balls to sort (the constant mode corresponding to $l=0$ belongs exclusively to the Neumann case)
\begin{equation}
    \mbox{deg}_N(l)=\frac{(l+n-1)!}{l!\,(n-1)!}, \qquad\qquad
\mbox{deg}_D(l)=\frac{(l+n-2)!}{(l-1)!\,(n-1)!}.
\end{equation}
The degeneracy for the full sphere is then the sum of both degeneracies. We can address now the multiplicative anomaly on spheres between two generic linear factors by setting all quasi-periods to one and considering arguments $\frac{n-1}{2}+a$ and $\frac{n-1}{2}+b$ for Neumann boundary condition and $1+\frac{n-1}{2}+a$ and $1+\frac{n-1}{2}+b$, for Dirichlet. With these building blocks, we can compute first the multiplicative anomaly to build up the shifted Laplacian $L_a$ by setting $b=-a$, i.e. $\mbox{MA}(D_a, D_{-a})$, and then the multiplicative anomaly among shifted Laplacians $L_a$ and $L_b$ based on the accumulative and associative properties of the defects:
\begin{eqnarray}\label{accum}
2\cdot\mbox{MA}(L_{a},L_{b})&=&\mbox{MA}(D_{a},D_{b})+\mbox{MA}(D_{a},D_{-b})+\mbox{MA}(D_{-a},D_{b})+\mbox{MA}(D_{-a},D_{-b})\nonumber\\\nonumber\\
&-&\mbox{MA}(D_{a},D_{-a})- \mbox{MA}(D_{b},D_{-b})~.
\end{eqnarray}

Let us mention in advance that the explicit results we will find follow the general rule that the multiplicative anomaly for Neumann and Dirichlet boundary conditions happen to be opposite in sign in odd dimensions, adding up to zero, whereas in even dimensions they are identical. The values for the multiplicative anomaly between linear factors that build up the shifted Laplacians coincide with those reported by Dowker using spectral techniques and expanding the zeta functions in terms of the shift (cf.~\cite{Dowker:2014tea}, eqn.31\footnote{We are grateful to J.S. Dowker for his help in fixing few numerical coefficients and signs in a previous version of this paper.}), and the same holds between shifted Laplacian (cf.~\cite{Dowker:2014tea}, eqn.15). Interestingly, for even spheres the same multiplicative anomaly between the linear factors in the shifted Laplacian had been previously obtained in~\cite{Cognola_2015} via Wodzicki residue and, as noticed in \cite{Dowker:2014xca}, also obtained in~\cite{Denef:2009kn} while computing the partition function for a massive scalar in (Euclidean) de Sitter space as a regularized product of quasinormal frequencies.

\subsection{Two-sphere:}
All quasi-periods set to one result in a quadratic polynomial in the arguments $(w,z)$
\begin{equation}
\mbox{MA}(A,B)=\frac{(w-z)^2}{4}~.
\end{equation}
The eigenvalues of the shifted Laplacian $L_a=-\nabla^2+1/4-a^2$ on the two-sphere are then obtained with $w=1/2+a,\,z=1/2-a$ and with $w=3/2+a,\,z=3/2-a$ for Neumann and Dirichlet boundary conditions on the Equator, respectively.
Both multiplicative anomalies turn out to be equal and the combined multiplicative anomaly between the linear factors $D_{\pm a}=\sqrt{-\nabla^2+1/4}\pm{a}$ that build up the shifted Laplacian is obtained
\begin{equation}
\mbox{MA}(D_a,D_{-a})=2\,a^2~.
\end{equation}
By contrast, the multiplicative anomaly between a pair of shifted Laplacians  $L_{a}$ and $L_{b}$ vanishes
\begin{equation}
\mbox{MA}(L_{a},L_{b})=0~.
\end{equation}

\subsection{Three-sphere:} The generalized Shintani-Mizuno formula with all quasi-periods set to one yields now a cubic polynomial in the arguments $(w,z)$
\begin{equation}
\mbox{MA}(A,B)=-\frac{(w-z)^2\cdot(w+z-3)}{8}~.
\end{equation}
The eigenvalues of the shifted Laplacian $L_a=-\nabla^2+1-a^2$ on the three-sphere are obtained with $w=1+a,\,z=1-a$ and with $w=2+a,\,z=2-a$ for Neumann and Dirichlet boundary conditions on the Equator, respectively.
The multiplicative anomaly for Neumann b.c.  turns out to be $a^2/2$ opposite to that for Dirichlet b.c.
\begin{equation}
\mbox{MA}(D_a,D_{-a})|_{_{Neu}}=-\mbox{MA}(D_a,D_{-a})|_{_{Dir}}=\frac{1}{2}a^2,
\end{equation}
so that the combined multiplicative anomaly between the linear factors $D_{\pm a}=\sqrt{-\nabla^2+1}\pm{a}$ that build up the shifted Laplacian vanishes.

For the multiplicative anomaly between a pair of shifted Laplacians  $L_{a}$ and $L_{b}$ we again obtain vanishing results
\begin{equation}
\mbox{MA}(L_{a},L_{b})|_{_{\textrm{Neu}}}=\mbox{MA}(L_{a},L_{b})|_{_{\textrm{Dir}}}=0~.
\end{equation}

\subsection{Four-sphere:} With all quasi-periods set to one, the generalized Shintani-Mizuno formula produces a quartic polynomial in the arguments $(w,z)$
\begin{equation}
\mbox{MA}(A,B)=\frac{(w-z)^2\cdot(11w^2+14wz+11z^2-72w-72z+132)}{288}~.
\end{equation}
The eigenvalues of the shifted Laplacian $L_a=-\nabla^2+9/4-a^2$ on the four-sphere are then obtained with $w=3/2+a,\,z=3/2-a$ and with $w=5/2+a,\,z=5/2-a$ for Neumann and Dirichlet boundary conditions on the Equator, respectively.
Both multiplicative anomalies turn out to be equal and the combined multiplicative anomaly between the linear factors $D_{\pm a}=\sqrt{-\nabla^2+9/4}\pm{a}$ that build up the shifted Laplacian is obtained
\begin{equation}
\mbox{MA}(D_a,D_{-a})=\frac{2}{9}a^4\,-\,\frac{1}{12}a^2~.
\end{equation}
The multiplicative anomaly between a pair of shifted Laplacians  $L_{a}$ and $L_{b}$ is nontrivial now

\begin{equation}
\mbox{MA}(L_{a},L_{b})=\frac{(a^2-b^2)^2}{24}~.
\end{equation}

\subsection{Five-sphere:} All quasi-periods set to one in the generalized Shintani-Mizuno formula result now in a quintic polynomial in the arguments $(w,z)$
\begin{equation}
\mbox{MA}(A,B)=-\frac{(w-z)^2\cdot(w+z-5)\cdot(5w^2 + 2wz+ 5z^2 - 30w- 30z + 60)}{576}~.
\end{equation}
The eigenvalues of the shifted Laplacian $L_a=-\nabla^2+4-a^2$ on the three-sphere are obtained with $w=2+a,\,z=2-a$ and with $w=3+a,\,z=3-a$ for Neumann and Dirichlet boundary conditions on the Equator, respectively.

The multiplicative anomaly for Neumann boundary conditions turns out to be the opposite of that for Dirichlet, namely
\begin{equation}
\mbox{MA}(D_a,D_{-a})|_{_{Neu}}=-\mbox{MA}(D_a,D_{-a})|_{_{Dir}}= \frac{1}{18} a^{4}-\frac{1}{12} a^{2}.
\end{equation}
Therefore, the combined multiplicative anomaly between the linear factors $D_{\pm a}=\sqrt{-\nabla^2+4}\pm{a}$, that build up the shifted Laplacian, vanishes.\\
For the multiplicative anomaly between a pair of shifted Laplacians  $L_{a}$ and $L_{b}$ we obtain

\begin{equation}
\mbox{MA}(L_{a},L_{b})|_{_{\textrm{Neu}}}=-\mbox{MA}(L_{a},L_{b})|_{_{\textrm{Dir}}}=\frac{(a^2-b^2)^2}{96}~.
\end{equation}

\subsection{Six-sphere:} With all quasi-periods set to one, the generalized Shintani-Mizuno formula produces here a sextic polynomial in the arguments $(w,z)$
\begin{eqnarray}
  \mbox{MA}(A,B)  &=& \frac{(w-z)^2}{86400} \left(137w^4 + 202w^3z + 222w^2z^2 + 202wz^3 + 137z^4 \right. \nonumber \\
  &-& 2250w^3 - 3150w^2z - 3150wz^2 - 2250z^3 + 14025w^2 + 17850wz \nonumber \\
  &+& \left.14025z^2 - 40500w - 40500z + 49320\right)
\end{eqnarray}
The eigenvalues of the shifted Laplacian $L_a=-\nabla^2+25/4-a^2$ on the six-sphere are then obtained with $w=5/2+a,\,z=5/2-a$ and with $w=7/2+a,\,z=7/2-a$ for Neumann and Dirichlet boundary conditions on the Equator, respectively.
Both multiplicative anomalies turn out to be equal and the combined multiplicative anomaly between the linear factors $D_{\pm a}=\sqrt{-\nabla^2+25/4}\pm{a}$ that build up the shifted Laplacian is obtained
\begin{equation}
\mbox{MA}(D_a,D_{-a})=\frac{23\,}{2700}a^6\,-\,\frac{1}{36} a^4\,+\,\frac{3}{320} a^2.
\end{equation}
The multiplicative anomaly between a pair of shifted Laplacians  $L_{a}$ and $L_{b}$
is then
\begin{equation}
\mbox{MA}(L_{a},L_{b})=\frac{(a^2-b^2)^2\cdot(2a^2+2b^2-5)}{960}~.
\end{equation}

\section{Examples: shifted conformal Laplacian on $S^1_{\beta}\times S^{n-1}$}
Let us now consider a temperature circle times the round sphere.
This time, the conformal Laplacian $Y=-\nabla^2+\frac{n-2}{4(n-1)}\, R=-\partial_0^2-\vec{\nabla}^2+\frac{(n-2)^2}{4}\equiv -\partial_0^2+\Delta_0$ factorizes into linear factors provided one of the quasi-periods is purely imaginary, say $\tau=\frac{2\pi\, i}{\beta}$. Again, considering the Neumann and Dirichlet problems on the $(n-1)$-sphere one can trade the orbital number $l$ by $n-1$ non-negative integers $m_1,m_2,...,m_{n-1}$. The same can be done with the winding number on the temperature circle, introducing an additional counting number $m_n$ and Neumann and Dirichlet boundary conditions on the circle.\\
We compute first for generic arguments $(w,z)$ and then restrict them to get the four combinations of boundary conditions Neumann-Neumann, Neumann-Dirichlet, Dirichlet-Neumann and Dirichlet-Dirichlet.\\
A quite surprising fact will be evident from the examples below: \\

\fbox{\parbox[b][15mm][c]{0.9\linewidth}{The standard Casimir energy for the Laplacian and for shifted Laplacians turns out to be exactly equal to the multiplicative anomaly among the linear factors that build them up~!}}

\subsection{Two-torus}
The linear factors in this case are $(m_1+m_2\tau+w)$ and $(m_1+m_2\overline{\tau}+z)$, with $\Im{(\tau)}>0$. The generalized Shintani-Mizuno formula for the multiplicative anomaly produces an overall factor of $i\pi$ from the logarithm of the ratio of $\tau$ and $\overline{\tau}=-\tau$, accompanied by a quadratic polynomial in the arguments $(w,z)$. The remarkable feature of the outcome is that it turns out to be linear in the inverse temperature $\beta$, just as the vacuum or Casimir energy:
\begin{equation}
\mbox{MA}(A,B)=-\beta\cdot \frac{(z + w)\cdot(z + w-2)}{16}-\beta\cdot \frac{1}
{24}~.
\end{equation}
The multiplicative anomaly among the linear factors that build up the conformal Laplacian $Y=D\cdot\overline{D}$ can then be worked out as the sum of the four contributions with $(w,z)$ equal to $(0,0), (1,1),(\tau,\overline{\tau})$ and  $(1+\tau,1+\overline{\tau})$ for N-N, N-D, D-N and D-D boundary conditions, respectively,
\begin{equation}
\mbox{MA}(D,\overline{D})= -\beta\cdot \frac{1}{6}~.
\end{equation}
More generally, let us allow for a shift in each of the linear factors $D+a$ and $\overline{D}+b$
\begin{equation}
\label{MA2D}
\mbox{MA}(D+a,\overline{D}+b)= -\beta\cdot \frac{(a+b)^2}{4}-\beta\cdot \frac{1}{6}~.
\end{equation}
There are now two alternative ways to build up shifted conformal Laplacians depending on where the shift is located, on the spatial (sphere) part or the temperature (circle) part.
For the spatial shift $Y_a=-\partial_0^2+(\sqrt{\Delta_0}+a)^2=(D+a)\cdot(\overline{D}+a)$
\begin{equation}
\mbox{MA}(D+a,\overline{D}+a)= -\beta\cdot a^2-\beta\cdot \frac{1}{6} ,
\end{equation}
whereas for the temperature shift $K_a=-(i\sqrt{-\partial_0^2}+a)^2+\Delta_0=(D+a)\cdot(\overline{D}-a)$
\begin{equation}
\mbox{MA}(D+a,\overline{D}-a)= -\beta\cdot \frac{1}{6}~.
\end{equation}
One can also compute the multiplicative anomaly among shifted conformal Laplacians by exploiting the accumulative and associative properties (eqn.\ref{accum})
\begin{equation}
\label{torus-spatial}
\mbox{MA}(Y_a, Y_b)=\beta\cdot \frac{(a-b)^2}{4}~,
\end{equation}
and
\begin{equation}
\label{torus-temp4}
\mbox{MA}(K_a, K_b)= -\beta\cdot \frac{(a-b)^2}{4}~.
\end{equation}
Interestingly, the multiplicative anomaly among shifted Laplacians for the particular choice $b=-a$
\begin{equation}
\label{torus-temp2}
\mbox{MA}(Y_a,
Y_{-a})=-\mbox{MA}(K_a, K_{-a})= \beta\cdot a^2
\end{equation}
coincides with the multiplicative anomaly computed by Elizalde et al. via Wodzicki residue for free massless scalars, provided one identifies the combination charge times chemical potential with the shift $e\mu=a$ and compactifies the spatial direction to a circle $V_1=2\pi$(cf. eqns. 86 and 91 in~\cite{Elizalde:1997sw}).

But let us return to the multiplicative anomaly for $Y_a=-\partial_0^2+(\sqrt{\Delta_0}+a)^2$. This linear term in $\beta$ will enter the partition function and contribute to the large-$\beta$ asymptotics determining the Casimir energy. The multiplicative anomaly turns up then in the exponential with an additional factor of $-\frac{1}{2}$ and should be compared with the leading behavior dominated by the vacuum or Casimir energy $-\beta\, E_0$. It is immediately apparent that both are exactly equal
\begin{equation}
\label{CasSpatial-2D}
E_0=\frac{1}{2\,\beta}\,\mbox{MA}(D+a,\overline{D}+a)= -\frac{a^2}{2}-\frac{1}{12} ~.
\end{equation}
The standard value for $E_0$ is well known (see, e.g.~\cite{CAPPELLI1989707}). It can easily be computed in terms of Hurwitz zetas and their relation with Bernoulli polynomials
\begin{eqnarray}
E_0&=&\frac{1}{2}\left(\sum_{l=0}^{\infty}(l+a)+\sum_{l=1}^{\infty}(l+a)\right)=\zeta_H(-1,a)-\frac{a}{2}\nonumber\\\nonumber\\&=&-\frac{B_2(a)}{2}-\frac{a}{2}\,=\,-\frac{a^2}{2}-\frac{1}{12}~.
\end{eqnarray}
The same happens for the temperature-shifted Laplacian $K_a$. The partition functions turn out to be dominated by the vacuum energy
\begin{equation}
\label{CasTemp-2D}
E_0=\frac{1}{2\,\beta}\,\mbox{MA}(D+a,\overline{D}-a)=-\frac{1}{12} ~.
\end{equation}

\subsection{$S^1_{\beta} \times S^3$}
The linear factors now are $(m_1+m_2+m_3+m_4\tau+w)$ and $(m_1+m_2+m_3+m_4\overline{\tau}+z)$. The formula for the multiplicative anomaly produces a quartic polynomial in the arguments $(w,z)$:
\begin{equation}
\mbox{MA}(A,B)=-\beta\cdot\frac{  \left(z+w\right)^2 \left(z+w-6\right)^2}{768} -\beta\cdot\frac{ \left(z+w\right) \left(z+w-6\right)}{64}-\beta\cdot\frac{19}{480}
\end{equation}
The multiplicative anomaly among the linear factors that build up the conformal Laplacian $Y=D\cdot\overline{D}$ can now be worked out as the sum of the four contributions with $(w,z)$ equal to $(1,1), (2,2),(1+\tau,1+\overline{\tau})$ and  $(2+\tau,2+\overline{\tau})$ for N-N, N-D, D-N, and D-D boundary conditions, respectively,
\begin{equation}
\mbox{MA}(D,\overline{D})= \beta\cdot \frac{1}{120}~.
\end{equation}

Allowing for a shift in each of the linear factors $D+a$ and $\overline{D}+b$, we get
\begin{equation}
\label{MA4D}
\mbox{MA}(D+a,\overline{D}+b)= -\beta\cdot\frac{\left(a+b\right)^{4}}{192} + \beta\cdot\frac{1}{120}~.
\end{equation}

For the spatial shift in the conformal Laplacian $Y_a=-\partial_0^2+(\sqrt{\Delta_0}+a)^2=(D+a)\cdot(\overline{D}+a)$
\begin{equation}
\mbox{MA}(D+a,\overline{D}+a)= -\beta\cdot \frac{a^4}{12}+\beta\cdot \frac{1}{120}~,
\end{equation}
whereas for the temperature shift $K_a=-(i\sqrt{-\partial_0^2}+a)^2+\Delta_0=(D+a)\cdot(\overline{D}-a)$
\begin{equation}
\mbox{MA}(D+a,\overline{D}-a)= \beta\cdot \frac{1}{120}~.
\end{equation}
Exploiting the accumulative properties (eqn.\ref{accum}), the multiplicative anomaly among shifted conformal Laplacians turns out to be \begin{equation}
\label{S1S3-spatial}
\mbox{MA}(Y_a, Y_b)=\beta\cdot \frac{(a-b)^2\,(7a^2+10ab+7b^2)}{192}
~,
\end{equation}
and
\begin{equation}
\label{S1S3-temp}
\mbox{MA}(K_a, K_b)= -\beta\cdot \frac{(a-b)^4}{192}~.
\end{equation}
Here, we again notice that the multiplicative anomaly among shifted Laplacians for the particular choice $b=-a$
\begin{equation}
\mbox{MA}(Y_a, Y_{-a})=-\mbox{MA}(K_a, K_{-a})= \beta\cdot \frac{a^4}{12}
\end{equation}
coincides with the multiplicative anomaly computed by Elizalde et al. via Wodzicki residue for free massless scalars, provided the spatial directions are compactified to the three-sphere $V_3=2\pi^2$ (cf. eqns. 88 and 94 in~\cite{Elizalde:1997sw}).

Going back to the multiplicative anomaly for $Y_a=-\partial_0^2+(\sqrt{\Delta_0}+a)^2$, we verify again the equality with the Casimir energy
\begin{equation}
\label{CasSpatial-4D}
E_0=\frac{1}{2\,\beta}\,\mbox{MA}(D+a,\overline{D}+a)= -\frac{a^4}{24}+\frac{1}{240} ~.
\end{equation}
The standard value for $E_0$ (see, e.g. \cite{Gibbons:2006ij}) can again be computed in terms of Hurwitz zetas and their relation with Bernoulli polynomials. The degeneracy $(l+1)^2$ needs to be expanded in powers of $(l+1+a)$
\begin{eqnarray}
2\,E_0&=&\sum_{l=0}^{\infty}(l+1)^2\,(l+1+a)=\zeta_H(-3,1+a)-2a\,\zeta_H(-2,1+a)+a^2\,\zeta_H(-1,1+a)\nonumber\\\nonumber\\&=&-\frac{B_4(-a)}{4}-2a\,\frac{B_3(-a)}{3}-a^2\,\frac{B_2(-a)}{2}\,=\,-\frac{a^4}{12}+\frac{1}{120}~.
\end{eqnarray}
The same happens for the temperature-shifted Laplacian $K_a$. The partition functions turn out to be dominated by the vacuum energy
\begin{equation}
\label{CasTemp-4D}
E_0=\frac{1}{2\,\beta}\,\mbox{MA}(D+a,\overline{D}-a)=\frac{1}{240} ~.
\end{equation}

\subsection{$S^1_{\beta} \times S^5$}
In this case, we need to add two more counters on the sphere to the linear factors, $(m_1+m_2+m_3+m_4+m_5+m_6\tau+w)$ and $(m_1+m_2+m_3+m_4+m_5+m_6\overline{\tau}+z)$, and we end up with a sextic  polynomial in the arguments $(w,z)$ for the multiplicative anomaly:
\begin{eqnarray}
    \mbox{MA}(A,B) &=&-\beta\cdot\frac{\! \left(z+w\right)^{3} \left(z-10+w\right)^{3}}{92160} \nonumber \\
    &-&\beta\cdot\frac{5 \left(z+w\right)^{2} \left(z-10+w\right)^{2}}{9216} \nonumber \\
    &-& \beta\cdot\frac{19 \left(z+w\right) \left(z-10+w\right)}{2304} \nonumber \\
    &-&\beta\cdot\frac{863}{24192} \label{MABS1S5}
\end{eqnarray}
To compute the multiplicative anomaly among the linear factors that build up the conformal Laplacian $Y=D\cdot\overline{D}$ there are four contributions with $(w,z)$ equal to $(2,2), (3,3),(2+\tau,2+\overline{\tau})$ and  $(3+\tau,3+\overline{\tau})$ coming from N-N, N-D, D-N, and D-D boundary conditions, respectively,
\begin{equation}
\mbox{MA}(D,\overline{D})= -\beta\cdot \frac{31}{30240}~.
\end{equation}
Allowing again for a shift in each of the linear factors $D+a$ and $\overline{D}+b$, we get
\begin{equation}
\label{MA6D}
\mbox{MA}(D+a,\overline{D}+b)= -\frac{\beta  \left(a+b\right)^{6}}{23040}+\frac{\beta  \left(a+b\right)^{4}}{2304}-\frac{31 \beta}{30240}.
\end{equation}
Now, for the spatial shift in the conformal Laplacian $Y_a=-\partial_0^2+(\sqrt{\Delta_0}+a)^2=(D+a)\cdot(\overline{D}+a)$
\begin{equation}
\mbox{MA}(D+a,\overline{D}+a)=-\beta\cdot\frac{84 a^{6}-210 a^{4}+31}{30240}~,
\end{equation}
whereas for the temperature shift $K_a=-(i\sqrt{-\partial_0^2}+a)^2+\Delta_0=(D+a)\cdot(\overline{D}-a)$
\begin{equation}
\mbox{MA}(D+a,\overline{D}-a)= -\beta\cdot \frac{31}{30240}~.
\end{equation}
Exploiting the accumulative and associative properties (eqn.\ref{accum}), the multiplicative anomaly among shifted conformal Laplacians turns out to be
\begin{equation}
\label{S1S5-spatial}
\mbox{MA}(Y_a, Y_b)=\frac{\beta  \left(a-b\right)^{2} \left(31 a^{4}+56 a^{3} b+66 a^{2} b^{2}+56 a \,b^{3}+31 b^{4}-70 a^{2}-100 a b-70 b^{2}\right)}{23040},
\end{equation}
and
\begin{equation}
\label{S1S5-temp}
\mbox{MA}(K_a, K_b)=-\frac{\beta  \left(a^{2}-2 a b+b^{2}-10\right) \left(a-b\right)^{4}}{23040}~.
\end{equation}
As in lower dimensions, for the particular choice $b=-a$ the multiplicative anomaly among shifted Laplacians
\begin{equation}
\mbox{MA}(Y_a, Y_{-a})=-\mbox{MA}(K_a, K_{-a})= \frac{\beta  \left(2a^{2}-5\right) a^4}{720}
\end{equation}
may be compared with the multiplicative anomaly computed by Elizalde et al. via Wodzicki residue for free massless scalars, provided the spatial directions are compactified to the five-sphere $V_5=\pi^3$ (cf. eqns. 96 and 97 in~\cite{Elizalde:1997sw}). Curiously, the agreement is now only achieved for the leading power.

Going back to the multiplicative anomaly for $Y_a=-\partial_0^2+(\sqrt{\Delta_0}+a)^2$, we can verify again the equality with the Casimir energy
\begin{equation}
\label{CasSpatial-6D}
E_0=\frac{1}{2\,\beta}\,\mbox{MA}(D+a,\overline{D}+a)= -\frac{84 a^{6}-210 a^{4}+31}{60480}~.
\end{equation}
The standard value for $E_0$ (see, e.g. \cite{Gibbons:2006ij}) can again be computed in terms of Hurwitz zetas and their relation with Bernoulli polynomials, this time through a lengthier calculation. The degeneracy $\frac{(n+1)(n+2)^2(n+3)}{12}$ needs to be expanded in powers of $(n+2+a)$
\begin{eqnarray}
2\,E_0&=&\sum_{n=0}^{\infty}\frac{(n+1\,)(n+2)^2\,(n+3)}{12}\,(n+2+a)
\nonumber\\\nonumber\\&=&\frac{1}{12}\,\zeta_H(-5,2+a)-\frac{1}{3} a\cdot\zeta_H(-4,2+a)+\frac{6a^{2}-1}{12}\cdot\zeta_H(-3,2+a)\nonumber\\\nonumber\\& &
-\frac{a(2a^2-1)}{6}\cdot\zeta_H(-2,2+a)+\frac{a^2(a^2-1)}{12}\cdot\zeta_H(-1,2+a)\nonumber\\\nonumber\\
&=&-\frac{84 a^{6}-210 a^{4}+31}{30240}~.
\end{eqnarray}
The same happens again for the temperature-shifted Laplacian $K_a$. The partition functions turn out to be dominated by the vacuum energy
\begin{equation}
\label{CasTemp-6D}
E_0=\frac{1}{2\,\beta}\,\mbox{MA}(D+a,\overline{D}-a)=-\frac{31}{60480}~.
\end{equation}

\section {On Shintani's proof of the Kronecker limit formula}
\label{Section 6}

Let us consider the zeta function
\begin{equation}
\xi(s,w,\tau)=\sum_{m,n\in \mathbf{Z}} |m+n\,\tau + w|^{-2s}~,
\end{equation}
with $\tau$ and $w$ complex, assuming $Im(\tau)>0$ and $m+n\,\tau+w\neq 0$ to avoid a null term\footnote{The case $w=\mathbf{Z}+\mathbf{Z}\,\tau$ can be readily obtained by carefully suppressing the zero factor in the final expression.}.
The version of the Kronecker (second) limit formula for the derivative with respect to $s$ at $s=0$, rather than for the value at $s=1$ (c.f.~\cite{Stark1977inbook}), as worked out by Shintani \cite{10.3836/tjm/1270472992} consists in the following closed expression for the $\zeta$-regularized product
\begin{equation}
\xi'(0,w,\tau) = -\log\left| \frac{\vartheta(w,\tau)}{\eta(\tau)} e^{\frac{i\pi w(w-\bar{w})}{\tau-\bar{\tau}}}\right|^2
\end{equation}
where Dedekind's eta and Jacobi's eta functions are given by
\begin{eqnarray}
    \eta(\tau) &=&  e^{i \pi \frac{\tau}{12}} \prod_{n-1}^{\infty} (1- e^{2\pi i \tau}), \\
    \vartheta(w,\tau) &=& 2 e^{i \pi  \frac{\tau}{6}} \sin(\pi w) \eta(\tau) \prod_{n=1}^{\infty} (1 - e^{2\pi i(n\tau+w)}) (1 - e^{2\pi i(n\tau-w)})~.
\end{eqnarray}

Shintani's approach proceeded by first splitting up the double sum
\begin{eqnarray}
\xi(s,w,\tau)&=&\sum_{m,n\geq 0} \left\{\,|m+n\,\tau + w|^{-2s}+|m-n\,\tau +1-w|^{-2s}\right.\nonumber\\\nonumber\\
& & \left.+|m-n\,\tau + w-\tau|^{-2s}+|m+n\,\tau +1-w+\tau|^{-2s}\right\}~,
\end{eqnarray}
 followed by splitting the regularized products on each term `liberating' the Barnes' gamma factors and paying the price of the multiplicative anomaly\footnote{The relevant anomaly is $\mbox{MA}(D_{(w,\tau)}, D_{(z,\overline{\tau})})= \frac{\tau-\overline{\tau}}{4\tau\,\overline{\tau}}\cdot B_2(\frac{\tau\,z-\overline{\tau}\,w}{\tau-\overline{\tau}})\cdot(\log\tau-\log\overline{\tau} )$. The appropriate log-branch for opposite quasi-periods $-\tau$ and $-\overline{\tau}$ requires $\log(-\tau)=-i\pi+\log(\tau)$ and $\log(-\overline{\tau})=i\pi+\log(\overline{\tau})$, respectively.}
 \begin{eqnarray}
-\xi'(0,w,\tau)&=&-\log |\Gamma_2(w|1,\tau)\,\Gamma_2(1-w|1,-\tau)\,\Gamma_2(w-\tau|1,-\tau)\,\Gamma_2(1-w+\tau|1,\tau)|^2 \nonumber\\\nonumber\\
&&+ \mbox{MA}(D_{(w,\tau)},D_{(\overline{w},\overline{\tau})})+ \mbox{MA}(D_{(1-w,-\tau)},D_{(1-\overline{w},-\overline{\tau})})\nonumber\\\nonumber\\
&&+ \mbox{MA}(D_{(w-\tau,-\tau)},D_{(\overline{w}-\overline{\tau},-\overline{\tau})})+ \mbox{MA}(D_{(1-w+\tau,\tau)},D_{(1-\overline{w}+\overline{\tau},\overline{\tau})})
\nonumber\\\nonumber\\
&=&-\log |\Gamma_2(w|1,\tau)\,\Gamma_2(1-w|1,-\tau)\,\Gamma_2(w-\tau|1,-\tau)\,\Gamma_2(1-w+\tau|1,\tau)|^2 \nonumber\\\nonumber\\
&&+\,i\pi\frac{\tau-\overline{\tau}}{\tau\,\overline{\tau}}\cdot B_2(\frac{\tau\,\overline{w}-\overline{\tau}\,w}{\tau-\overline{\tau}})
~.
\end{eqnarray}
The reflection formula for Barnes double gamma (see, e.g. proposition 6.1 in \cite{Friedman2004362}) came into play here to further reduce to infinite convergent products (further recast in terms of Jacobi theta and Dedekind eta functions)
\begin{eqnarray}
&&-\log |\Gamma_2(w|1,\tau)\,\Gamma_2(1-w|1,-\tau)\,\Gamma_2(w-\tau|1,-\tau)\,\Gamma_2(1-w+\tau|1,\tau)|^2 \nonumber \\
&& \nonumber \\
&&= \log \prod_{m\geq0}|(1-e^{2\pi\,i(w+m\tau)})(1-e^{2\pi\,i(\tau -w+m\tau)}) |^2 \nonumber \\
&& \nonumber \\
&& + \left\{i\pi\zeta_2(0,w|1,\tau)+i\pi\zeta_2(0,1-w+\tau|1,\tau)+ c.c.\right\}.
\end{eqnarray}

Let us now examine the consequence of  having chosen a different splitting, locating the $m=0$ term of the initial sum in the second and fourth terms
\begin{eqnarray}
\xi(s,w,\tau)&=&\sum_{m,n\geq 0} \left\{\,|m+n\,\tau + 1+w|^{-2s}+|m-n\,\tau-w|^{-2s}\right.\nonumber\\\nonumber\\
& & \left.+|m-n\,\tau +1+w-\tau|^{-2s}+|m+n\,\tau-w+\tau|^{-2s}\right\}~.
\end{eqnarray}
For the $\zeta$-regularized product, after `liberating' the Barnes' gamma factors and paying the price of the multiplicative anomaly, we now obtain
 \begin{eqnarray}
-\xi'(0,w,\tau)&=&-\log |\Gamma_2(1+w|1,\tau)\,\Gamma_2(-w|1,-\tau)\,\Gamma_2(1+w-\tau|1,-\tau)\,\Gamma_2(-w+\tau|1,\tau)|^2 \nonumber\\\nonumber\\
&&+ \mbox{MA}(D_{(1+w,\tau)},D_{(1+\overline{w},\overline{\tau})})+ \mbox{MA}(D_{(-w,-\tau)},D_{(-\overline{w},-\overline{\tau})})\nonumber\\\nonumber\\
&&+ \mbox{MA}(D_{(1+w-\tau,-\tau)},D_{(1+\overline{w}-\overline{\tau},-\overline{\tau})})+ \mbox{MA}(D_{(-w+\tau,\tau)},D_{(-\overline{w}+\overline{\tau},\overline{\tau})})
\nonumber\\\nonumber\\
&=&-\log |\Gamma_2(1+w|1,\tau)\,\Gamma_2(-w|1,-\tau)\,\Gamma_2(1+w-\tau|1,-\tau)\,\Gamma_2(-w+\tau|1,\tau)|^2 \nonumber\\\nonumber\\
&&+\,i\pi\frac{\tau-\overline{\tau}}{\tau\,\overline{\tau}}\cdot B_2(-\frac{\tau\,\overline{w}-\overline{\tau}\,w}{\tau-\overline{\tau}})
~.
\end{eqnarray}
Notice the subtle difference in the multiplicative anomaly, the argument of the Bernoulli polynomial comes out with the opposite sign. The reflection formula for Barnes double gamma produces the very same  infinite convergent products (further recast in terms of Jacobi theta and Dedekind eta functions) but different zeta prefactors
\begin{eqnarray}
&&-\log |\Gamma_2(1+w|1,\tau)\,\Gamma_2(-w|1,-\tau)\,\Gamma_2(1+w-\tau|1,-\tau)\,\Gamma_2(-w+\tau|1,\tau)|^2 \nonumber\\
&& \nonumber \\
&&= \log \prod_{m\geq0}|(1-e^{2\pi\,i(w+m\tau)})(1-e^{2\pi\,i(\tau -w+m\tau)}) |^2 \nonumber \\
&& \nonumber \\
&& + \left\{i\pi\zeta_2(0,1+w|1,\tau)+i\pi\zeta_2(0,-w+\tau|1,\tau)+ c.c.\right\}.
\end{eqnarray}

Had we been a little cavalier concerning the multiplicative anomaly and not included it in the first place, we would then have had a discrepancy for the $\zeta$-regularized products depending on the initial splitting\footnote{This is very reminiscent of the two different prescriptions in the `one-step' regularization of the infinite products with two complex quasi-periods of~\cite{Cabo-Bizet:2018ehj}. In our particular case, the possible discrepancy in the final answer for any choice of the splitting is cured by the multiplicative anomaly.}. The apparent discrepancy would be the difference between the Barnes zeta terms\footnote{The difference is easily computed due to the recurrence relation for Barnes multiple zetas (cf. eqn.1.2 in \cite{RUIJSENAARS2000107})
$
\zeta_2(0,w|1,\tau)-\zeta_2(0,1+w|1,\tau)=\zeta_1(0,w|\tau)=\frac{1}{2}-\frac{w}{\tau}
$.} in the exponentials:

\begin{eqnarray}
&&\left\{i\pi\zeta_2(0,w|1,\tau)+i\pi\zeta_2(0,1-w+\tau|1,\tau)+ c.c.\right\}\nonumber\\\nonumber\\
&-&\left\{i\pi\zeta_2(0,1+w|1,\tau)+i\pi\zeta_2(0,-w+\tau|1,\tau)+ c.c.\right\}~\nonumber\\\nonumber\\
&=&-2 i\pi\left\{\frac{w}{\tau}-\frac{\overline{w}}{\overline{\tau}}\right\}~.
\end{eqnarray}
However, by taking into consideration the additional term given by the multiplicative anomaly we have an additional contribution
\begin{eqnarray}
&&\,i\pi\frac{\tau-\overline{\tau}}{\tau\,\overline{\tau}}\cdot B_2(\frac{\tau\,\overline{w}-\overline{\tau}\,w}{\tau-\overline{\tau}})-\,i\pi\frac{\tau-\overline{\tau}}{\tau\,\overline{\tau}}\cdot B_2(-\frac{\tau\,\overline{w}-\overline{\tau}\,w}{\tau-\overline{\tau}})\nonumber\\\nonumber\\
&=& 2 i\pi\left\{\frac{w}{\tau}-\frac{\overline{w}}{\overline{\tau}}\right\}~,
\end{eqnarray}
that exactly cancels the mismatch and yields a unique answer for the $\zeta$-regularized product, i.e., the Kronecker second limit formula. In all, one can say that what saves the day is precisely the role of the multiplicative anomaly.

\section{Application: Casimir energy for GJMS operators}
It is known that there are two alternative factorizations of the GJMS operators on $S^1_{\beta}\times S^{n-1}$ in terms of shifted conformal Laplacian (see, e.g.,~\cite{juhl2010conformally, Beccaria_2016, Beccaria:2017dmw}), given by
\begin{equation}
P_{2k}=\prod_{j=1}^{k}\left\{-\partial_o^{\,2}\,+\,(\sqrt{\Delta_0}+2j-k-1)^2\right\}=\prod_{j=1}^{k}\left\{-(i\sqrt{-\partial_o^2}+2j-k-1)^{\,2}\,+\,\Delta_0\right\}
\end{equation}
It is worth noticing that the factorization of the eigenvalues into linear factors is unique, but two different pairings lead to the two alternative quadratic factorizations into shifted conformal Laplacians.

The conventional computation of the one-loop partition function, or functional determinant, yields different results for the Casimir energy under $\zeta$-regularization. However, in this section, we will show their equivalence once the multiplicative anomaly is properly taken into account.

Let us first compute the accumulated Casimir energy for the shifted conformal Laplacian factors and then add up the corresponding multiplicative anomaly among them. The latter is given by the averaged multiplicative anomaly between all possible pairings by the pairwise-accumulative property.
\subsection{Two-torus}

\paragraph{Factorization with spatial shift:}

The standard Casimir energy for the GJMS operator is simply the sum of the individual ones (eqn.\ref{CasSpatial-2D})
\begin{equation}
E^{(k)}_0=\sum_{j=1}^k -\frac{6(2j-k-1)^2+1}{12}=
-\frac{1}{6}k^3+ \frac{1}{12}k~.
\end{equation}
Notice the conflict for $k>1$ with the universal relation for a two-dimensional CFT where $E_0=-\frac{c}{12}$, since the central charge for the GJMS operators in 2D is $k^3$ (see, e.g.~\cite{Diaz:2008hy,Dowker:2010bu,Dowker:2010qy}).

Let us include now the correction to the Casimir energy coming from the multiplicative anomaly between the shifted conformal Laplacian. The multiplicative anomaly, being pairwise accumulative, equals the average among all pairs
\begin{equation}
\frac{1}{k}\sum_{1\leq j,l\leq k}\mbox{MA}(Y_{(2j-k-1)},
 Y_{(2l-k-1)})~.
\end{equation}
Plugging in eqn.\ref{torus-spatial}, we get
\begin{equation}
\mbox{MA}=\beta\cdot\left(\frac{1}{6}k^3-\frac{1}{6}k\right).
\end{equation}
For the GJMS operator, the improved Casimir energy becomes
\begin{equation}
\tilde{E}^{(k)}_0=E^{(k)}_0+\frac{1}{2\beta}\mbox{MA}=-\frac{k^3}{12}~,
\end{equation}
restoring the universality.

\paragraph{Factorization with temperature shift:}

For the alternative factorization, the standard Casimir energy for the GJMS operator is again the sum of the individual ones (eqn.\ref{CasTemp-2D})
\begin{equation}
E^{(k)}_0=\sum_{j=1}^k \left(-\frac{1}{12}\right)=-\frac{1}{12}k~.
\end{equation}
Again the result is in conflict for $k>1$ with the expectation for a $CFT_2$.
Including now the corrections to the Casimir energy coming from the multiplicative anomaly between the shifted conformal Laplacian
\begin{equation}
\frac{1}{k}\sum_{1\leq j,l\leq k}\mbox{MA}(K_{2j-k-1)}, K_{(2l-k-1)})~,
\end{equation}
and plugging in eqn.\ref{torus-temp4}, we get instead
\begin{equation}
\mbox{MA}=-\beta\cdot\left(\frac{1}{6}k^3-\frac{1}{6}k\right).
\end{equation}
For the GJMS operator, the improved Casimir energy becomes
\begin{equation}
\tilde{E}^{(k)}_0=E^{(k)}_0+\frac{1}{2\beta}\mbox{MA}=-\frac{k^3}{12},
\end{equation}
restoring the universality and the agreement between the two factorizations.\\
We find out another remarkable fact, readily verified in this case by using eqn.\ref{MA2D},

\begin{equation}
\label{MAamongAll}
\tilde{E}^{(k)}_0=\frac{1}{2\beta}\frac{1}{k}\sum_{1\leq j,l\leq k}\mbox{MA}(D_{(2j-k-1)}, \overline{D}_{(2l-k-1)})~.
\end{equation}

\fbox{\parbox[b][15mm][c]{0.9\linewidth}{The common value for the improved Casimir energy can also be obtained as the multiplicative anomaly among all linear factors that build up the GJMS operator, this decomposition being unique.}}

\subsection{$S_{\beta}^{1} \times S^3$}

\paragraph{Factorization with spatial shift:}

The standard Casimir energy for the GJMS operator (cf.~\cite{Beccaria:2017dmw}) is simply given by the sum of the individual ones (eqn.\ref{CasSpatial-4D})
\begin{equation}
E^{(k)}_0=\sum_{j=1}^k -\frac{10 \, (2j-k-1)^{4}-1}{240}=-\frac{k \left(6 k^{4}-20 k^{2}+11\right)}{720}~.
\end{equation}
We now include the corrections to the Casimir energy from the multiplicative anomaly between the shifted conformal Laplacian. The multiplicative anomaly, being the average among all pairs, equals
\begin{equation}
\frac{1}{k}\sum_{1\leq j,l\leq k}\mbox{MA}(Y_{(2j-k-1)},
 Y_{(2l-k-1)})~.
\end{equation}
Plugging in eqn.\ref{S1S3-spatial}, we obtain
\begin{equation}
\mbox{MA}=\beta\cdot\frac{k \left(k^2-1\right) \left(4 k^{2}-11\right) }{360}.
\end{equation}
For the GJMS operator, the improved Casimir energy becomes
\begin{equation}
\tilde{E}^{(k)}_0=E^{(k)}_0+\frac{1}{2\beta}\mbox{MA}=-\frac{k^{3} \left(2 k^{2}-5\right)}{720}~.
\end{equation}

\paragraph{Factorization with temperature shift:}

For the alternative factorization, the standard Casimir energy for the GJMS operator is again the sum of the individual ones (eqn.\ref{CasTemp-4D})
\begin{equation}
E^{(k)}_0=\sum_{j=1}^k \frac{1}{240}=\frac{k}{240}~.
\end{equation}
We now include the corrections to the Casimir energy coming from the multiplicative anomaly between the shifted conformal Laplacian
\begin{equation}
\frac{1}{k}\sum_{1\leq j,l\leq k}\mbox{MA}(K_{(2j-k-1)}, K_{(2l-k-1)})~.
\end{equation}
Plugging in eqn.\ref{S1S3-temp}, we obtain instead
\begin{equation}
\mbox{MA}=-\beta\frac{k  \left(k^2-1\right) \left(2 k^{2}-3\right) }{360}.
\end{equation}
For the GJMS operator, the improved Casimir energy then becomes
\begin{equation}
\tilde{E}^{(k)}_0=E^{(k)}_0+\frac{1}{2\beta}\mbox{MA}=-\frac{k^{3} \left(2 k^{2}-5\right)}{720}~,
\end{equation}
attaining agreement between the two factorizations. Again this common value for the improved Casimir energy can also be obtained as the multiplicative anomaly (eqn.\ref{MA4D}) among all linear factors that build up the GJMS operator.

 To discuss yet another feature of this improved Casimir energy, we make a brief digression here. In a four-dimensional CFT, according to Cappelli and Coste~\cite{CAPPELLI1989707}, the Casimir energy is related to the coefficients of the trace anomaly
 \begin{equation}
     E_o=\frac{3}{4}\left(a+\frac{1}{2}g\right)~,
 \end{equation}
where $a$ is the type-A central charge and $g$ is the coefficient of the total derivative in
\begin{equation}
    (4\pi)^2 \langle T\rangle = -a\,E_4 + c\,W^2 + g\,\nabla^2 R~.
\end{equation}
The total derivative term is what makes the Casimir energy scheme dependent. For free conformal fields, the trace anomaly can be read off from the heat kernel coefficients.
For example, sticking to zeta-regularization, for the conformal Laplacian one finds
\begin{equation}
    [\,a\,,\,c\,,\,g\,]=[\,\frac{1}{360}\,,\,\frac{1}{120}\,,\,\frac{1}{180}\,]
\end{equation}
and the Casimir energy $E_0=\frac{1}{240}$.
For the Paneitz operator, in turn,
\begin{equation}
    [\,a\,,\,c\,,\,g\,]=[\,-\frac{7}{90}\,,\,-\frac{1}{15}\,,\,\frac{1}{15}\,]
\end{equation}
the standard Casimir energy $E_o^{(2)}=-3/40$ fails to comply with the Capelli-Coste relation, whereas it surprisingly holds for the improved Casimir energy $\tilde{E}_o^{(2)}=-1/30$. Unfortunately, even the first few heat coefficients for higher-derivative operators remain largely unknown, and total derivative terms are usually discarded. One notable exception is Branson's computation for the Paneitz operator~\cite{Branson96}~\footnote{Actually, he reported for the heat coefficient $[\,(c-a)/2\,,\,-2\,a\,,\,-3g+2a\,]=[\,1/4\,,\,7\,,\,-16\,]/45$ in a basis where he traded the Euler density by his Q-curvature, which contains itself also a total derivative (cf. Lemma 2 and the subsequent evaluation at $m=4$ in~\cite{Branson96}). To compare with the trace anomaly, the heat coefficient must be multiplied by two because of the quartic nature of the Paneitz operator.}, from where we extracted the value $g=1/15$ above\footnote{The same value can also be worked out from the expression found by Gusynin~\cite{Gusynin:1989ky} for quartic operators. We are grateful to L. Casarin for bringing this paper to our attention.}. Our prediction then is that the coefficient of the total derivative term in the heat kernel coefficient for GJMS operators is the one related to the improved Casimir energy via the Cappelli-Coste relation.

\subsection{$S_{\beta}^{1} \times S^5$}

\paragraph{Factorization with spatial shift:}
The standard Casimir energy for the GJMS operator is given by the sum of the individual ones (eqn.\ref{CasSpatial-6D})
\begin{equation}
E^{(k)}_0=\sum_{j=1}^k -\frac{84 (2j-k-1)^{6}-210 (2j-k-1)^{4}+31}{60480}=-\frac{k \left(12 k^{6}-126 k^{4}+336 k^{2}-191\right)}{60480}~.
\end{equation}
Let us include now the corrections to the Casimir energy coming from the multiplicative anomaly between the shifted conformal Laplacian. The multiplicative anomaly, being pairwise accumulative, equals the average among all pairs
\begin{equation}
\frac{1}{k}\sum_{1\leq j,l\leq k}\mbox{MA}(Y_{(2j-k-1)},
 Y_{(2l-k-1)})~.
\end{equation}
Plugging in eqn.\ref{S1S5-spatial}, we get
\begin{equation}
\mbox{MA}=\beta\cdot\frac{k \left(k^2-1\right) \left(9 k^{4}-89 k^{2}+191\right)} {30240}~.
\end{equation}
The improved Casimir energy for the GJMS operator then becomes
\begin{equation}
\tilde{E}^{(k)}_0=E^{(k)}_0+\frac{1}{2\beta}\mbox{MA}=-\frac{k^{3} \left(3 k^{4}-28 k^{2}+56\right)}{60480}~.
\end{equation}

\paragraph{Factorization with temperature shift:}

For the alternative factorization, the standard Casimir energy for the GJMS operator is again the sum of the individual ones (eqn.\ref{CasTemp-6D})
\begin{equation}
E^{(k)}_0=\sum_{j=1}^k -\frac{31 }{60480}=-\frac{31 k}{60480}~.
\end{equation}
We now include the corrections to the Casimir energy coming from the multiplicative anomaly between the shifted conformal Laplacian
\begin{equation}
\frac{1}{k}\sum_{1\leq j,l\leq k}\mbox{MA}(K_{2j-k-1)}, K_{(2l-k-1)})~.
\end{equation}
Plugging in eqn.\ref{S1S5-temp}, we get instead
\begin{equation}
\mbox{MA}=-\frac{k \left(k^2-1\right) \left(3 k^{4}-25 k^{2}+31\right) \beta}{30240}~.\end{equation}
For the GJMS operator, the improved Casimir energy becomes
\begin{equation}
\tilde{E}^{(k)}_0=E^{(k)}_0+\frac{1}{2\beta}\mbox{MA}=-\frac{k^{3} \left(3 k^{4}-28 k^{2}+56\right)}{60480}~,
\end{equation}
achieving the agreement between the two factorizations. We stress again that this common value for the improved Casimir energy can also be obtained as the multiplicative anomaly (eqn.\ref{MA6D}) among all linear factors that build up the GJMS operator.

\section{Summary and outlook}
We have succeeded in extending the Shintani-Mizuno expression for the multiplicative anomaly of linear factors and used it to reproduce known results for Laplacians on spheres. Regarding thermal partition functions for different factorizations of higher-derivative operators, we have shown they agree once the multiplicative anomaly is properly included. This yields a modified (\textit{improved}) Casimir energy that dominates the zero temperature limit. In addition, we have found out that the standard Casimir energy for (shifted) Laplacians precisely coincides with the multiplicative anomaly among the linear factors \footnote{Remarkably, this role of the multiplicative anomaly can already be appreciated in Kronecker limit formula and the determinant of the Laplacian on the torus, as stressed in~\cite{Quine1993}.
The product of the determinants of the linear factors differs from that of the Laplacian by a multiplicative anomaly, although not captured by Wodzicki's formula.}. For GJMS operators, the improved Casimir energy restores the universal relation with the central charge in two dimensions, whereas in four dimensions it reconciles with the Cappelli-Coste relation for the Paneitz operator.
Although established for the case of scalar Laplacians and their conformal powers (GJMS operators), this may well hold for Laplacians and higher-derivative operators on vector, tensor, and even higher-spin fields.  \\
Regarding the ambiguity of the Casimir energy in four (and higher even) dimensions, it can be traced back to local finite counterterms which are the conformal primitives of the trivial total derivatives or trivial anomalies in the trace anomaly. On the conformally flat $S^1_{\beta}\times S^{n-1}$ backgrounds, the universal part of the Casimir energy that depends on the type-A central charge is already known~\cite{Herzog:2013ed}, but this is only valid in a particular regularization scheme where all trivial divergences in the trace anomaly are discarded. This scheme certainly differs from $\zeta$ regularization, which produces a particular combination of trivial total derivatives. In 4D the ambiguity is controlled by the coefficient $g$ of $\nabla^2R$ in the trace anomaly, as shown by Capelli and Coste~\cite{CAPPELLI1989707}
\begin{equation}
    E_o=\frac{3}{4}a+\frac{3}{8}g~.
\end{equation}
In 6D things are more complicated, there is a basis of six independent trivial anomalies~\cite{Bastianelli:2000rs} and the universal part obtained by Herzog and Huang~\cite{Herzog:2013ed} must be supplemented by the coefficients of these trivial total derivatives.
Prompted by the result of Cappelli and Coste in 4D, we have obtained the following extension to 6D (further details\footnote{We have verified the validity of this expression in all 6D cases considered in~\cite{Bastianelli:2000hi}, where the coefficients $g$'s were computed via heat kernel, against the Casimir energies computed in~\cite{Gibbons:2006ij}.} will be given elsewhere~\cite{Upcomming})
\begin{equation}
    E_o=-\frac{15}{8}a-\frac{5}{12}\left(g_5+\frac{1}{4}g_7+\frac{1}{2}g_8-10g_9+g_{10}\right)~,
\end{equation}
where $a$ is the 6D type-A trace anomaly coefficient and the $g$'s are the coefficients of the six independent trivial anomalies $M_5, M_6, M_7, M_8, M_9$ and $M_{10}$ of~\cite{Bastianelli:2000rs}. Alternatively, in a 6D conformally flat background, the above basis is redundant and one can simplify further to get, in terms of the Schouten scalar $J$ and the Schouten tensor $V$, Branson's basis (see, e.g.~\cite{BransonSharp1995}) for trivial total derivatives  $\nabla^2\nabla^2 J$, $\nabla^2J^2$ and $\nabla^2|V|^2$ with coefficients $\gamma_1, \gamma_2$ and $\gamma_3$, respectively,    
\begin{equation}
    E_o=-\frac{15}{8}a-\frac{1}{192}\left(8\gamma_1-8\gamma_2 +11\gamma_3\right)~.
\end{equation}
The matching we have found between the multiplicative anomaly and Casimir energy in 4D and 6D holds whenever the Casimir energy is computed in $\zeta$ regularization and the trivial total derivative coefficients are obtained as well via heat heat kernel in $\zeta$ regularization. 
For GJMS operators, in particular, the heat kernel computation should produce coefficients g's that match the improved Casimir energy. This claim remains a prediction for other than the conformal Laplacian or Yamabe operator, except for the Paneitz operator in 4D where the explicit coefficients have been worked out and the matching, via Cappelli-Coste relation, was successfully verified. 
\\
As for the physical interpretation, the Casimir energy in 4D and 6D remains ambiguous due to the above-mentioned trivial total derivatives terms in the trace anomaly. We emphasize that the equivalence we found applies to a particular regularization scheme ($\zeta$ regularization), so that the inclusion of the multiplicative anomaly can be traced back to the addition of a precise combination of finite local counterterms. There is certainly no new physics in the inclusion of the multiplicative anomaly; however, if one sticks to $\zeta$ regularization then consistency and conformity with trivial total derivatives, regardless factorization choices, demands a proper account of the multiplicative anomaly.    
\\
There are several instances where the role of the multiplicative anomaly seems worth to be revisited. A prominent example is the supersymmetric version of the Casimir energy \cite{Assel:2015nca} that ought to be \textit{physical} and connected with the central charges of the CFT. A multiplicative anomaly might turn up in the traditional manipulation of one-loop functional determinants, as shown in Shintani's derivation of the Kronecker limit formula (Section~\ref{Section 6}) and the example in \ref{Addendum2},  as well as with the inclusion of higher-derivative multiplets~\cite{Beccaria:2018rxp}.\\
Finally, it seems natural to ask whether the multiplicative anomaly and its connection with the CFT Casimir energy may find its place in a dual holographic counterpart.

\ack
We thank F. Bastianelli, L. Casarin, J.S. Dowker, A. Monin, and especially E. Friedman for valuable conversations and comments. We are also grateful to the anonymous referee for helpful suggestions and clarifications. This work was partially funded through FONDECYT-Chile 1220335.
D.E.D. wishes to salute Harald Dorn and Hans-J\"org Otto on the occasion of the 30th anniversary of the DOZZ formula.
\appendix
\section{Bernoulli polynomials of higher degree}

Let us write down the explicit form of the first few Bernoulli polynomials of higher degree that enter the integral formula.
The generating function
\begin{equation}
\frac{t^n\,e^{-w\,t}}{\prod_{i=1}^n\left\{1-e^{-a_i\,t}\right\}}=\sum_{l=0}^{\infty}B_{n,l}(w| \vec{a})\,\frac{t^l}{l!},
\end{equation}
determines the polynomial $B_{n,l}(w| \vec{a})$. The explicit expressions up to order five are the following
\begin{eqnarray}
B_{1,1}(w|a)&=&\frac{1}{2}-\frac{w}{\sigma_1}~,\\
B_{2,2}(w|a_1,a_2)&=&\frac{\sigma_1^2+\sigma_2}{6\sigma_2}-\frac{\sigma_1}{\sigma_2}\,w+\frac{w^2}{\sigma_2}~,\\
B_{3,3}(w|a_1,a_2,a_3)&=&\frac{\sigma_1\sigma_2}{4\sigma_3}-\frac{\sigma_1^2+\sigma_2}{2\sigma_3}\,w+\frac{3\sigma_1}{2\sigma_3}\,w^2-\frac{w^3}{\sigma_3},\\
B_{4,4}(w|a_1,a_2,a_3,a_4)&=&\frac{4\sigma_1^2\sigma_2+3\sigma_2^2-\sigma_1^4+\sigma_1\sigma_3-\sigma_4}{30\sigma_4}-\frac{\sigma_1\sigma_2}{\sigma_4}\,w+\frac{\sigma_1^2+\sigma_2}{\sigma_4}\,w^2 \nonumber \\
&-&\frac{2\sigma_1}{\sigma_4}\,w^3+\frac{w^4}{\sigma_4}~,
\\
B_{5,5}(w|a_1,a_2,a_3,a_4,a_5)&=&-\frac{\sigma_1(\sigma_1^2\sigma_2-3\sigma_2^2+\sigma_1\sigma_3-\sigma_4)}{12\sigma_5} \nonumber \\
&-& \frac{4\sigma_1^2\sigma_2+3\sigma_2^2-\sigma_1^4+\sigma_1\sigma_3-\sigma_4}{6\sigma_5}\,w \nonumber \\
&+&\frac{5\sigma_1\sigma_2}{2\sigma_5}\,w^2- 5 \frac{\sigma_1^2+\sigma_2}{3\sigma_5}\,w^3+\frac{5\sigma_1}{2\sigma_5}\,w^4-\frac{w^5}{\sigma_5}~.
\end{eqnarray}
where the $\sigma's$ are the elementary symmetric functions
\begin{equation}
\sigma_k=\sum_{1\leq r_1<r_2< \ldots r_k\leq n}a_{r_1}a_{r_2}\ldots a_{r_k}.
\end{equation}

\section{Barnes multiple gammas and reflection formulas}\label{Addendum2}

Let us try to emulate Shintani's derivation of the Kronecker limit formula in the case of two quasi-complex periods and examine the role of the multiplicative anomaly.
Consider the zeta function
\begin{equation}
\xi(s,w|\tau,\sigma)=\sum_{m\in \mathbf{Z};n,l\geq0} |m+n\,\tau + l\,\sigma+w|^{-2s}~,
\end{equation}
with $\tau, \sigma$ and $w$ complex, assuming $\Im(\tau)>0, \Im(\sigma)>0$ and $m+n\,\tau+ l\,\sigma+w\neq 0$ to avoid a null term.
Splitting up the sum over integer $m$ as
\begin{equation}
\xi(s,w|\tau,\sigma)=\sum_{m,n,l\geq 0} \left\{\,|m+n\,\tau+l\,\sigma + w|^{-2s}+|m-n\,\tau-l\,\sigma + 1-w|^{-2s}\right\},
\end{equation}
 and hence the regularized products can be written in terms of Barnes gamma factors by paying the price of the multiplicative anomaly\footnote{The multiplicative anomaly $\mbox{MA}(D_{(w,\tau,\sigma)},D_{(\overline{w},\overline{\tau},\overline{\sigma})})$, as computed with the generalized Shintami-Mizuno formula, is given by $\frac{\tau(\overline{w}-\frac{\overline{\sigma}+1}{2})-(w-\frac{\sigma+1}{2})\overline{\tau}}{6\tau\overline{\tau}(\tau-\overline{\tau})(\tau\overline{\sigma}-\sigma\overline{\tau})}\left\{[\tau(\overline{w}-\frac{\overline{\sigma}+1}{2})-(w-\frac{\sigma+1}{2})\overline{\tau}]^2-\frac{(\tau\overline{\sigma}-\sigma\overline{\tau})^2}{4}-\frac{(\tau-\overline{\tau})^2}{4}\right\}(\log\tau -\log\overline{\tau})\,+\,\left\{\binom{\tau}{\overline{\tau}}\leftrightarrow\binom{\sigma}{\overline{\sigma}}\right\}$}
 \begin{eqnarray}
-\xi'(0,w|\tau,\sigma)&=&-\log |\Gamma_3(w|1,\tau,\sigma)\,\Gamma_3(1-w|1,-\tau,-\sigma)|^2 \nonumber\\\nonumber\\
&&+ \mbox{MA}(D_{(w,\tau,\sigma)},D_{(\overline{w},\overline{\tau},\overline{\sigma})})+ \mbox{MA}(D_{(1-w,-\tau,-\sigma)},D_{(1-\overline{w},-\overline{\tau},-\overline{\sigma})})\nonumber\\\nonumber\\
&=&-\log |\Gamma_3(w|1,\tau,\sigma)\,\Gamma_3(1-w|1,-\tau,-\sigma)|^2 \nonumber\\\nonumber\\
&&-2\pi i\,\frac{\tau(\overline{w}-\frac{\overline{\sigma}+1}{2})-(w-\frac{\sigma+1}{2})\overline{\tau}}{6\tau\overline{\tau}(\tau-\overline{\tau})(\tau\overline{\sigma}-\sigma\overline{\tau})}\left\{\left[\tau(\overline{w}-\frac{\overline{\sigma}+1}{2})-(w-\frac{\sigma+1}{2})\overline{\tau}\right]^2\right.\nonumber\\
& &\nonumber\\
& &\left.-\frac{(\tau\overline{\sigma}-\sigma\overline{\tau})^2}{4}-\frac{(\tau-\overline{\tau})^2}{4}\right\} +\,\left\{\binom{\tau}{\overline{\tau}}\leftrightarrow\binom{\sigma}{\overline{\sigma}}\right\}.
\end{eqnarray}
The reflection formula for Barnes double gamma (see, e.g. proposition 6.1 in \cite{Friedman2004362}) comes into play here to further reduce to infinite convergent products
\begin{eqnarray}
&&-\log |\Gamma_3(w|1,\tau,\sigma)\,\Gamma_3(1-w|1,-\tau,-\sigma)|^2 \\\nonumber\\
&&= \log \prod_{m,n\geq0}|1-e^{2\pi\,i(w+m\tau+n\sigma)}|^2 + \left\{i\pi\zeta_3(0,w|1,\tau\sigma)+ c.c.\right\}~.\nonumber
\end{eqnarray}
Let us now examine the consequence of  having chosen a different splitting, locating the $m=0$ term of the initial sum in the second term
\begin{eqnarray}
\xi(s,w|\tau,\sigma)&=&\sum_{m,n,l\geq 0} \left\{\,|m+n\,\tau+l\,\sigma + 1+w|^{-2s}+|m-n\,\tau-l\,\sigma-w|^{-2s}\right\}~.
\end{eqnarray}
For the $\zeta$-regularized product, after `liberating' the Barnes' gamma factors and paying the price of the multiplicative anomaly, we obtain now
 \begin{eqnarray}
-\xi'(0,w|\tau,\sigma)&=&-\log |\Gamma_3(1+w|1,\tau,\sigma)\,\Gamma_3(-w|1,-\tau,-\sigma)|^2 \nonumber\\\nonumber\\
&&+ \mbox{MA}(D_{(1+w,\tau,\sigma)},D_{(1+\overline{w},\overline{\tau},\overline{\sigma})})+ \mbox{MA}(D_{(-w,-\tau,-\sigma)},D_{(-\overline{w},-\overline{\tau},-\overline{\sigma})})\nonumber\\\nonumber\\
&=&-\log |\Gamma_3(1+w|1,\tau,\sigma)\,\Gamma_3(-w|1,-\tau,-\sigma)|^2 \nonumber\\\nonumber\\
&&-2\pi i\,\frac{\tau(\overline{w}-\frac{\overline{\sigma}-1}{2})-(w-\frac{\sigma-1}{2})\overline{\tau}}{6\tau\overline{\tau}(\tau-\overline{\tau})(\tau\overline{\sigma}-\sigma\overline{\tau})}\left\{\left[\tau(\overline{w}-\frac{\overline{\sigma}-1}{2})-(w-\frac{\sigma-1}{2})\overline{\tau}\right]^2\right.\nonumber\\
& &\nonumber\\
&-&\left.\frac{(\tau\overline{\sigma}-\sigma\overline{\tau})^2}{4}-\frac{(\tau-\overline{\tau})^2}{4}\right\}+\,\left\{\binom{\tau}{\overline{\tau}}\leftrightarrow\binom{\sigma}{\overline{\sigma}}\right\}
\end{eqnarray}

Notice the subtle difference in the multiplicative anomaly. The reflection formula for Barnes' double-$\Gamma$ produces the very same infinite convergent products, but different exponential prefactors
\begin{eqnarray}
&&-\log |\Gamma_3(1+w|1,\tau,\sigma)\,\Gamma_3(-w|1,-\tau,-\sigma)|^2 \\\nonumber\\
&&= \log \prod_{m,n\geq0}|1-e^{2\pi\,i(w+m\tau+n\sigma)}|^2 + \left\{i\pi\zeta_3(0,1+w|1,\tau,\sigma)+ c.c.\right\}~.\nonumber
\end{eqnarray}
The apparent discrepancy would be the difference between the Barnes zeta terms\footnote{The difference is again easily computed due to the recurrence relation for Barnes multiple $\zeta$'s (cf. eqn.1.2 in \cite{RUIJSENAARS2000107})
$
\zeta_3(0,w|1,\tau,\sigma)-\zeta_3(0,1+w|1,\tau,\sigma)=\zeta_2(0,w|\tau,\sigma)=\frac{w^2}{2\tau\sigma}-\frac{\tau+\sigma}{2\tau\sigma}w+\frac{\tau^2+\sigma^2+3\tau\sigma}{12\tau\sigma}
$.} in the exponentials:
\begin{eqnarray}
&&\left\{i\pi\zeta_3(0,w|1,\tau,\sigma)+ c.c.\right\}-\left\{i\pi\zeta_3(0,1+w|1,\tau,\sigma)+ c.c.\right\}\nonumber\\\nonumber\\
&=&i\pi\left\{\frac{w^2}{2\tau\sigma}-\frac{\overline{w}^2}{2\overline{\tau}\overline{\sigma}} -\frac{\tau+\sigma}{2\tau\sigma}w+\frac{\overline{\tau}+\overline{\sigma}}{2\overline{\tau}\overline{\sigma}}\overline{w}+\frac{\tau^2+\sigma^2}{12\tau\sigma}-\frac{\overline{\tau}^2+\overline{\sigma}^2}{12\overline{\tau}\overline{\sigma}}\right\}~.
\end{eqnarray}
Nonetheless, by taking into consideration the additional term given by the multiplicative anomaly we have an additional contribution that exactly cancels the mismatch and yields a unique answer for the $\zeta$-regularized product.

\section*{References}


\begin{thebibliography}{10}

\bibitem{RAY1971145}
D.~Ray and I.~Singer, {\it R-torsion and the Laplacian on Riemannian
  manifolds},  {\em Advances in Mathematics} {\bf 7} (1971), no.~2 145--210.

\bibitem{Kontsevich1995}
M.~Kontsevich and S.~Vishik, {\em Geometry of determinants of elliptic
  operators}, pp.~173--197.
\newblock Birkh{\"a}user Boston, Boston, MA, 1995.

\bibitem{wodzicki1987noncommutative}
M.~Wodzicki, {\it Noncommutative residue},  {\em K-Theory, Arithmetic and
  Geometry} (1987) 320.

\bibitem{Wodzicki1987}
M.~Wodzicki, {\em Noncommutative residue Chapter I. Fundamentals},
  pp.~320--399.
\newblock Springer Berlin Heidelberg, Berlin, Heidelberg, 1987.

\bibitem{GUILLEMIN1985131}
V.~Guillemin, {\it A new proof of Weyl's formula on the asymptotic distribution
  of eigenvalues},  {\em Advances in Mathematics} {\bf 55} (1985), no.~2
  131--160.

\bibitem{ShintaniOnValues}
T.~Shintani, {\it On values at $s=1$ of certain l functions of totally real
  algebraic number fields},  in {\em Proceedings of the Taniguchi International
  Symposium, Division of Mathematics} (S.~Iyanaga, ed.), vol.~1 of {\em
  Algebraic Number Theory}, pp.~201--212, Japan Society for the Promotion of
  Science, 1977.

\bibitem{10.3836/tjm/1270472992}
T.~Shintani, {\it {A Proof of the Classical Kronecker Limit Formula}},  {\em
  Tokyo Journal of Mathematics} {\bf 3} (1980), no.~2 191 -- 199.

\bibitem{barnes1904theory}
E.~W. Barnes, {\it On the theory of the multiple gamma function},  {\em Trans.
  Cambridge Philos. Soc.} {\bf 19} (1904) 374--425.

\bibitem{Friedman2004362}
E.~Friedman and S.~Ruijsenaars, {\it Shintani-Barnes zeta and gamma functions},
   {\em Advances in Mathematics} {\bf 187} (2004), no.~2 362 -- 395.

\bibitem{MIZUNO2006155}
Y.~Mizuno, {\it Generalized {Lerch} formulas: Examples of zeta-regularized
  products},  {\em Journal of Number Theory} {\bf 118} (2006), no.~2 155--171.

\bibitem{CastilloFriedman2011}
V.~Castillo-Garate and E.~Friedman, {\it Discrepancies of products of
  zeta-regularized products},  {\em Mathematical Research Letters} {\bf 19}
  (07, 2012) 199--212.

\bibitem{CastilloFriedman2012}
V.~Castillo-Garate, E.~Friedman, and M.~Mantoiu, {\it The multiplicative
  anomaly of three or more commuting elliptic operators},  {\em Mathematical
  Research Letters} {\bf 22} (11, 2012).

\bibitem{Dowker:2014tea}
J.~S. Dowker, {\it {Calculation of the multiplicative anomaly}},
  \href{http://xxx.lanl.gov/abs/1412.0549}{{\tt arXiv:1412.0549}}.

\bibitem{Beccaria:2014xda}
M.~Beccaria and A.~A. Tseytlin, {\it {Higher spins in AdS$_{5}$ at one loop:
  vacuum energy, boundary conformal anomalies and AdS/CFT}},  {\em JHEP} {\bf
  11} (2014) 114, [\href{http://xxx.lanl.gov/abs/1410.3273}{{\tt
  arXiv:1410.3273}}].

\bibitem{GJMSOriginal1992}
C.~R. Graham, R.~Jenne, L.~J. Mason, and G.~A.~J. Sparling, {\it Conformally
  invariant powers of the laplacian, i: Existence},  {\em Journal of the London
  Mathematical Society} {\bf s2-46} (1992), no.~3 557--565,
  [\href{http://xxx.lanl.gov/abs/https://londmathsoc.onlinelibrary.wiley.com/doi/pdf/10.1112/jlms/s2-46.3.557}{{\tt
  https://londmathsoc.onlinelibrary.wiley.com/doi/pdf/10.1112/jlms/s2-46.3.557}}].

\bibitem{BransonSharp1995}
T.~P. Branson, {\it Sharp inequalities, the functional determinant, and the
  complementary series},  {\em Transactions of the American Mathematical
  Society} {\bf 347} (1995), no.~10 3671--3742.

\bibitem{Gover2005}
A.~Gover, {\it Laplacian operators and $Q$-curvature on conformally Einstein
  manifolds},  {\em Mathematische Annalen} {\bf 336} (07, 2005).

\bibitem{juhl2010conformally}
A.~Juhl, {\it On conformally covariant powers of the Laplacian},
  \href{http://xxx.lanl.gov/abs/0905.3992}{{\tt arXiv:0905.3992}}.

\bibitem{Beccaria_2016}
M.~Beccaria and A.~A. Tseytlin, {\it Iterating free-field AdS/CFT: higher spin
  partition function relations},  {\em Journal of Physics A: Mathematical and
  Theoretical} {\bf 49} (Jun, 2016) 295401.

\bibitem{ErdelyBateman}
A.~Erd\'elyi, W.~Magnus, F.~Oberhettinger, and F.~Tricomi, {\em Higher
  Transcendental Functions}, vol.~I of {\em Bateman Manuscript Project}, ch.~I,
  p.~39.
\newblock Mc GrawHill Book Company, 1953.

\bibitem{RUIJSENAARS2000107}
S.~Ruijsenaars, {\it On Barnes' multiple zeta and gamma functions},  {\em
  Advances in Mathematics} {\bf 156} (2000), no.~1 107--132.

\bibitem{Dowkercmp/1104270298}
J.~S. Dowker, {\it {Effective action in spherical domains}},  {\em
  Communications in Mathematical Physics} {\bf 162} (1994), no.~3 633 -- 647.

\bibitem{Cognola_2015}
G.~Cognola, E.~Elizalde, and S.~Zerbini, {\it Functional determinant of the
  massive Laplace operator and the multiplicative anomaly},  {\em Journal of
  Physics A: Mathematical and Theoretical} {\bf 48} (Jan, 2015) 045203.

\bibitem{Dowker:2014xca}
J.~S. Dowker, {\it {Massive sphere determinants}},
  \href{http://xxx.lanl.gov/abs/1404.0986}{{\tt arXiv:1404.0986}}.

\bibitem{Denef:2009kn}
F.~Denef, S.~A. Hartnoll, and S.~Sachdev, {\it {Black hole determinants and
  quasinormal modes}},  \href{http://xxx.lanl.gov/abs/0908.2657}{{\tt
  arXiv:0908.2657}}.

\bibitem{Elizalde:1997sw}
E.~Elizalde, A.~Filippi, L.~Vanzo and S.~Zerbini, {\it One loop effective potential for a fixed charged self-interacting bosonic model at finite temperature with its related multiplicative anomaly},  {\em Phys. Rev. D} {\bf 57} (1998) 7430--7443.
[\href{http://xxx.lanl.gov/abs/hep-th/9710171}{{\tt hep-th/9710171}}].

 \bibitem{CAPPELLI1989707}
A.~Cappelli and A.~Coste, {\it On the stress tensor of conformal field theories
  in higher dimensions},  {\em Nuclear Physics B} {\bf 314} (1989), no.~3
  707--740.

\bibitem{Gibbons:2006ij}
G.~W. Gibbons, M.~J. Perry, and C.~N. Pope, {\it {Partition functions, the
  Bekenstein bound and temperature inversion in anti-de Sitter space and its
  conformal boundary}},  {\em Phys. Rev. D} {\bf 74} (2006) 084009,
  [\href{http://xxx.lanl.gov/abs/hep-th/0606186}{{\tt hep-th/0606186}}].

\bibitem{Stark1977inbook}
H.~Stark, {\em Class fields and modular forms of weight one}, pp.~277--287.
\newblock Lecture Notes in Mathematics.
\newblock Springer, 11, 2006.

\bibitem{Cabo-Bizet:2018ehj}
A.~Cabo-Bizet, D.~Cassani, D.~Martelli, and S.~Murthy, {\it {Microscopic origin
  of the Bekenstein-Hawking entropy of supersymmetric AdS$_{5}$ black holes}},
  {\em JHEP} {\bf 10} (2019) 062,
  [\href{http://xxx.lanl.gov/abs/1810.1144}{{\tt arXiv:1810.1144}}].

\bibitem{Beccaria:2017dmw}
M.~Beccaria and A.~A. Tseytlin, {\it {C$_{T}$ for higher derivative conformal
  fields and anomalies of (1, 0) superconformal 6d theories}},  {\em JHEP} {\bf
  06} (2017) 002, [\href{http://xxx.lanl.gov/abs/1705.0030}{{\tt
  arXiv:1705.0030}}].

\bibitem{Diaz:2008hy}
D.~E. Diaz, {\it {Polyakov formulas for GJMS operators from AdS/CFT}},  {\em
  JHEP} {\bf 07} (2008) 103, [\href{http://xxx.lanl.gov/abs/0803.0571}{{\tt
  arXiv:0803.0571}}].

\bibitem{Dowker:2010bu}
J.~Dowker, {\it {Entanglement entropy for even spheres}},
  \href{http://xxx.lanl.gov/abs/1009.3854}{{\tt arXiv:1009.3854}}.

\bibitem{Dowker:2010qy}
J.~Dowker, {\it {Determinants and conformal anomalies of GJMS operators on
  spheres}},  {\em J. Phys. A} {\bf 44} (2011) 115402,
  [\href{http://xxx.lanl.gov/abs/1010.0566}{{\tt arXiv:1010.0566}}].

\bibitem{Branson96}
T.~P. Branson,{\it {An anomaly associated with $4$-dimensional quantum gravity}},  {\em
  Communications in Mathematical Physics} {\bf 178} (1996), no.~2 301 -- 309.

\bibitem{Gusynin:1989ky}
V.~P.~Gusynin,
{\it {New Algorithm for Computing the Coefficients in the Heat Kernel Expansion}},
{\em Phys. Lett. B} \textbf{225} (1989), 233-239.


 \bibitem{Quine1993}
J.R.~Quine, S.H.~Heydari and R.Y.~Song, {\it Zeta regularized products},  {\em Trans. Am. Math. Soc.} \textbf{338} (2020) no.1, 213-231
(, Transactions of the American Mathematical Society, Vol. 338, No. 1, July 1993, pp. 213-231)

\bibitem{Assel:2015nca}
B.~Assel, D.~Cassani, L.~Di~Pietro, Z.~Komargodski, J.~Lorenzen, and
  D.~Martelli, {\it {The Casimir energy in curved space and its supersymmetric
  counterpart}},  {\em JHEP} {\bf 07} (2015) 043,
  [\href{http://xxx.lanl.gov/abs/1503.0553}{{\tt arXiv:1503.0553}}].

\bibitem{Beccaria:2018rxp}
M.~Beccaria and A.~A.~Tseytlin,
{\it {Superconformal index of higher derivative $\mathcal N=1$ multiplets in four dimensions}},
{\em JHEP} {\bf 10} (2018) 087,
[\href{http://xxx.lanl.gov/abs/1807.05911}{{\tt arXiv:1807.05911}}].

\bibitem{Herzog:2013ed}
C.~P.~Herzog and K.~W.~Huang,
{\it {Stress Tensors from Trace Anomalies in Conformal Field Theories}},
{\em Phys. Rev. D} \textbf{87} (2013), 081901
[arXiv:1301.5002 [hep-th]].

\bibitem{Bastianelli:2000rs}
F.~Bastianelli, G.~Cuoghi and L.~Nocetti,
\textit{ Consistency conditions and trace anomalies in six-dimensions},
{\em Class. Quant. Grav.} \textbf{18} (2001), 793-806
[arXiv:hep-th/0007222 [hep-th]].

\bibitem{Bastianelli:2000hi}
F.~Bastianelli, S.~Frolov and A.~A.~Tseytlin,
\textit{ Conformal anomaly of (2,0) tensor multiplet in six-dimensions and AdS / CFT correspondence},
{\em JHEP} \textbf{02} (2000), 013
[arXiv:hep-th/0001041 [hep-th]].

\bibitem{Gibbons:2006ij}
G.~W.~Gibbons, M.~J.~Perry and C.~N.~Pope,
\textit{ Partition functions, the Bekenstein bound and temperature inversion in anti-de Sitter space and its conformal boundary},{\em 
Phys. Rev. D} \textbf{74} (2006), 084009
[arXiv:hep-th/0606186 [hep-th]].

\bibitem{Upcomming}
R.~Aros, F.~Bugini and D.~E.~D\'iaz, \textit{ in preparation.}

\end{thebibliography}

\providecommand{\href}[2]{#2}\begingroup\raggedright\endgroup

\end{document}